\documentclass[nofootinbib,a4paper,prd,preprint,superscriptaddress,12pt,preprintnumbers,showpacs,tightenlines]{revtex4-1}

\usepackage{amsmath,amssymb,amsfonts}
\usepackage{graphicx,color}
\usepackage{bm}

\newcommand{\mpl}{M_{\rm Pl}}
\def\be{\begin{equation}}
\def\ee{\end{equation}}

\begin{document}


\title{Effect of transitions in the Planck mass during inflation\\ on
  primordial power spectra}
\author{Amjad Ashoorioon}
\email{A.Ashoorioon@lancaster.ac.uk}
\affiliation{Consortium for Fundamental Physics,
  Physics Department,\\
  Lancaster University, Lancaster, LA1 4YB,
  United Kingdom}
\author{Carsten van de Bruck}
\email{C.vandeBruck@sheffield.ac.uk}
\affiliation{Consortium for Fundamental Physics,
  School of Mathematics and Statistics,\\
  University of Sheffield, Hounsfield Road, Sheffield, S3 7RH,
  United Kingdom}
\author{Peter Millington}
\email{Peter.Millington@manchester.ac.uk}
\affiliation{Consortium for Fundamental Physics,
  School of Physics and Astronomy,\\
  University of Manchester, Oxford Road, Manchester, M13 9PL,
  United Kingdom}
\affiliation{Institute for Particle Physics Phenomenology,\\
  Durham University,
  Durham, DH1 3LE. United Kingdom}
\author{Susan Vu}
\email{Susan.Vu@sheffield.ac.uk}
\affiliation{Consortium for Fundamental Physics,
  School of Mathematics and Statistics,\\
  University of Sheffield, Hounsfield Road, Sheffield, S3 7RH,
  United Kingdom}

\preprint{MAN/HEP/2014/07, IPPP/14/59, DCPT/14/118}

\begin{abstract}
We study  the effect of sudden transitions  in the effective
Planck   mass   during   inflation   on  primordial   power   spectra.
Specifically, we consider models  in which this variation results from
the non-minimal coupling  of a Brans-Dicke type scalar  field. We find
that  the  scalar  power   spectra  develop  features  at  the  scales
corresponding to those leaving  the horizon during the transition.  In
addition, we  observe that the tensor perturbations  are {\it largely}
unaffected, so long  as the variation of the Planck  mass is below the
percent level.   Otherwise, the  tensor power spectra  exhibit damped
oscillations over the same scales.  Due to significant features in the
scalar power  spectra, the tensor-to-scalar ratio  $r$ shows variation
over the corresponding scales.  Thus, by studying the spectra of both
scalar and  tensor perturbations, one  can constrain sudden  but small
variations of  the Planck mass during inflation.   We illustrate these
effects with a  number of benchmark single- and  two-field models.  In
addition,  we comment  on their  implications and  the  possibility to
alleviate the tension between the observations of the tensor-to-scalar
ratio performed by the Planck and BICEP2 experiments.
\end{abstract}

\maketitle

\section{Introduction}
\label{intro}

Aside from resolving a number of issues in the standard
hot  Big  Bang  scenario (see  e.g.~\cite{Sarkar:1995dd,  Lyth:1998xn,
LiddleLyth,Baumann:2014nda}),  including  the  horizon problem  and  overabundance of  magnetic monopoles, inflationary cosmology
has made a number  of predictions consistent with current observations
of  the   Cosmic  Microwave   Background  (CMB).  These   include  the
expectation of an almost  spatially-flat universe and an approximately
scale-invariant   power   spectrum   of   the   primordial   curvature
perturbations,  as  confirmed by  the  Wilkinson Microwave  Anisotropy
Probe  (WMAP) \cite{Hinshaw:2012aka}  and, more  recently,  the Planck
Satellite \cite{Ade:2013uln}. 

Many models  of inflation also
predict a cosmological  gravitational wave background, parametrized in
terms of the tensor-to-scalar ratio $r$. Recently,   the  BICEP2   experiment  \cite{Ade:2014xna}   reported  a
detection of  B-mode polarization  in the CMB. When  interpreted as
being produced by such a background of primordial gravitational waves, this translates into a tensor-to-scalar ratio of
\be
  \label{bicep2r}
  r = 0.20^{+ 0.07}_{- 0.05}\;,
\ee
around   $\ell\simeq   80$,   i.e.~$k_{\rm   BICEP}\simeq   0.005~{\rm
Mpc}^{-1}$. This  would be in tension  with the previous  limit set by
Planck  \cite{Ade:2013uln} of $r\leq  0.11$ ($95\%$  C.L.) at  the pivot
scale $k_{0}\simeq 0.002~{\rm  Mpc}^{-1}$, i.e.~$\ell\simeq 28$.  Such
a value for  the amplitude of the gravitational  wave background would
determine the scale  of inflation to be at the  GUT scale. Moreover, a
value of  $r\geq 0.1$  would rule out  a large number  of inflationary
models in which  the displacement of the field  is smaller than $\mpl$
\cite{Lyth:1996im, Hotchkiss:2011gz, Antusch:2014cpa}.  Super-Planckian excursion of the
inflaton field  can be realized  within a number of  models, including
assisted inflation \cite{Liddle:1998jc, Kanti:1999vt, Copeland:1999cs, Mazumdar:2001mm, Jokinen:2004bp}, natural inflation \cite{Freese:1990rb, Adams:1992bn, Kim:2004rp}, string-inspired    many-field    models   \cite{    Dimopoulos:2005ac,
Ashoorioon:2009wa,Ashoorioon:2011ki, Ashoorioon:2014jja} and monodromy
\cite{Silverstein:2008sg,    McAllister:2014mpa,    Hebecker:2014eua}.
In addition, values of $r  >0.11$ can be realized
in  scalar   models  in  which  the  potential   has  flat  directions
\cite{Ballesteros:2014yva}.

Many authors  have proposed solutions to alleviate  the tension between BICEP2 and  Planck on both
theoretical  \cite{Wang:2014kqa, Ashoorioon:2014nta,  McDonald:2014kia, Kawasaki:2014fwa, Xu:2014laa, Bastero-Gil:2014oga,
 Cai:2014hja, Mukohyama:2014gba}     and    experimental     grounds    \cite{Liu:2014mpa,
Mortonson:2014bja,     Flauger:2014qra}. In non-singular bouncing cosmologies \cite{Mukhanov:1991zn}, it was shown \cite{Xia:2014tda} that the emergence of jump features in the scalar and tensor power spectra at a given scale may conspire to lessen this discrepancy. This tension can also be alleviated in Starobinsky models~\cite{Starobinsky:1992ts}, if the speed of the inflaton field undergoes a sudden change \cite{Contaldi:2014zua}. In addition, sharp features  are  observed in the scalar
power spectra in models of  punctuated inflation, where  the shape  of the  inflaton potential changes discontinuously at a given scale
\cite{Jain:2008dw, Jain:2009pm, Hazra:2010ve}. An inflaton  potential with a step was studied in the Einstein frame
\cite{Adams:2001vc} (see  also \cite{Joy:2007na}) and  shown to result
in oscillations in the power spectra. Fading oscillatory features in the primordial
scalar  power spectrum  can also  occur  from jumps  in the  potential
\cite{Ashoorioon:2006wc, Ashoorioon:2008qr,     Firouzjahi:2014fda,
Bartolo:2013exa,   Battefeld:2010rf},   particle   production   during
inflation  \cite{Barnaby:2010ke,  Battefeld:2010vr, Battefeld:2013bfl}
or turns in  the inflaton trajectory in the  landscape of heavy fields
\cite{Cespedes:2012hu,       Achucarro:2012fd,       Cespedes:2013rda,
Saito:2013aqa,Konieczka:2014zja}. For a discussion of observed features in the primordial power spectrum, see for instance \cite{Aslanyan:2014mqa} and references therein.

Recently, there has also been renewed interest in models of inflation with
time-varying  gravitational  constants \cite{Abolhasani:2012tz}.   The
potential  time-dependence   of  physical  constants   has  long  been
recognized   \cite{Dirac:1937ti}  and   variation  in   the  effective
gravitational  coupling is  known  to arise  in  theories of  modified
gravity,   such  as   Brans-Dicke   Scalar-Tensor  \cite{Brans:1961sx,
Fujii:1974bq,     Minkowski:1977aj,     Linde:1979kf,    Fujii:1982ms,
DeFelice:2011hq}     and    Tensor-Vector-Scalar     (TeVeS)    theory
\cite{Bekenstein:2004ne,  Bekenstein:2008pc}. The latter provides a
relativistic  basis for Milgrom's  Modified Newtonian  Dynamics (MOND)
\cite{Milgrom:1983ca}.     Time-dependency   may    also    arise   in
Supersymmetric  String   Theories  \cite{Wu:1986ac,  Ivashchuk:1988bs,
Ivashchuk:2014rga}.

The implications  of a  time-varying gravitational constant  (see also
\cite{Uzan:2010pm}) have been studied in the radiation and
matter  dominated epochs  \cite{Wetterich:1987fk}.  It was shown in \cite{Accetta:1990yb} that modulations of the    gravitational     constant can result from the non-minimal  coupling of a massive scalar field that
oscillates around  its vacuum expectation  value (VEV).  It  was found
that cosmological  measurements can be affected when  the frequency of
oscillations is high compared to the Hubble expansion rate.  This work
was  extended in  \cite{Steinhardt:1994vs} to consider a scenario in which the  Brans-Dicke field is driven  away from its VEV during inflation, thereby inducing oscillations. For cosmological perturbation theory in models beyond General Relativity, see \cite{Mukhanov:1990me}, \cite{Hwang:2005hb} and references therein.

There  are various  experiments that  test models  with a time-dependent
Newton's constant:  lunar ranging observations \cite{Williams:2004qba,
Muller:2007zzb},  Big  Bang  Nucleosynthesis (BBN)  \cite{Copi:2003xd,
Santiago:1997mu,   Bambi:2005fi,   Coc:2006rt},  gravitational   waves
\cite{Yunes:2009bv}  and,   more  recently,  the   WiggleZ  experiment
\cite{Nesseris:2011pc}.   In  addition, it  has  been  shown that  the
late-time evolution  of the gravitational constant  can be constrained
through comparisons of the  ages of globular clusters with independent
measurements of  the age of  the Universe \cite{Degl'Innocenti:1995nt}
and  by observations  of type  1a  supernovae \cite{GarciaBerro:2005yw,
Nesseris:2006jc},    as    well     as    pulsating    white    dwarfs
\cite{GarciaBerro:2011wc,           Corsico:2013ida},          pulsars
\cite{BisnovatyiKogan:2005ap,      Jofre:2006ug,      Verbiest:2008gy,
Lazaridis:2009kq}    and    neutron-star    surface    temperatures
\cite{Krastev:2007en}. Such limits on the variation of the gravitational
constant     place    constraints     on     scalar-tensor    theories
\cite{Babichev:2011iz} in addition to those obtained from observations
of     primordial     density    perturbations     \cite{Chiba:1997ij,
Tsujikawa:2000wc,  Lee:2010zy} and  gravitational  Cherenkov radiation
\cite{Kimura:2011qn}.

In this article, we show that sharp features may arise in the scalar power spectra as a result of transitions in the effective Planck mass (or equivalently Newton's  gravitational constant) during inflation. Specifically, we address the question of whether smooth
step variations  in the  gravitational coupling, occurring during the observable window of scales between 60 and 50 $e$-folds
before the end of inflation \cite{Turner:1993xz,Liddle:2003as}, have
sizeable  effects  on  the  power  spectra for  curvature  and  tensor
perturbations. Step changes in the Planck mass could result from a first-order phase
transition in the VEV of a Brans-Dicke field \cite{Abolhasani:2012tz}.
Alternatively,  as  we will  consider,  the  step  change could  arise
through a second-order  transition, with the Brans-Dicke field rolling
slowly towards its VEV.  We consider two scenarios: one in  which the role of Brans-Dicke field is
played by the inflaton itself and a two-field model in which this role
is played  by a second auxiliary  field. The variations  that  we have  in  mind  are  not violent  ones,  i.e.~the
variations of $\mpl$ are not of order $1$. Instead, they are typically
of order a percent or less. Nevertheless, we illustrate that, for a particular choice of parameters for the single-field model, the impact upon the resulting power spectra and, consequently, the tensor-to-scalar ratio $r$ has the potential to alleviate the aforementioned discrepancy between BICEP2 and Planck. Furthermore, in contrast to potentials with a step \cite{Adams:2001vc}, we
show that oscillations  are not observed in the scalar power spectra when the step transition occurs instead in the  non-minimal  coupling of  the  Brans-Dicke field.

The  paper  is  organized  as follows.   In  Sec.~\ref{sec:fe},  we
describe the relevant background  field and perturbation equations for
the      single-field     model      under      consideration.     In
Sec.~\ref{sec:model},  we solve  these systems  of equations numerically  for a
number  of single-  and two-field  benchmark models,  illustrating the
potential implications  for observations of  primordial power spectra.
Finally,  in Sec.~\ref{sec:con},  we provide  our  conclusions.  In
addition,   App.~\ref{sec:2f}   summarizes   the  background   and
perturbation  equations   for  the  two-field   model  considered  and
App.~\ref{sec:app1}  describes the approximate  analytic solutions
to the background evolution, relevant to Sec.~\ref{sec:model}.

\section{Field equations}
\label{sec:fe}

Our  goal is to  study the  influence of  variations in  the effective
Planck  mass  (or  equivalently  Newton's gravitational  constant)  on
inflation  and the  primordial  power spectra  for  scalar and  tensor
perturbations. This  can be  described  conveniently in the  context of
scalar-tensor  theories. Specifically, we  will focus  on the  case in
which the  evolution of the  inflaton itself causes this  variation by
means  of its  non-minimal  coupling  to the  Ricci  scalar. For  this
reason and throughout this article, we  choose to  perform the computations  in the  Jordan frame,
which  allows  us  to  model  these  variations  in  the  Planck  mass
intuitively.  Nevertheless, physical observables do not  depend on the  choice of frame
(see  e.g.~\cite{Kaiser:1994vs, Kaiser:1995nv, Prokopec:2013zya} and  references therein)  and hence
equivalent results would be obtained in the Einstein frame.

The single-field action that we consider is of the form
\be
\label{action}
  S^{(\varphi)}\ =\ \frac{1}{2} \int d^4 x \sqrt{-g}
      \bigg[\mpl^2 F(\varphi) R
         - g^{\mu \nu} \varphi_{,\mu} \varphi_{,\nu}
         - 2U(\varphi) \bigg]\;,
\ee
where $\mpl^2  = ( 8  \pi G_{\ast} )^{-1} $  is the reduced  Planck mass, 
with $G_*$ being the present-day Newton's constant; $R$ is the
Ricci scalar  and $U(\varphi)$  is the potential  of the  scalar field
$\varphi$.   Hereafter,  we  set  $\mpl  =1$,  with  all  dimensionful
quantities understood to  be in units of the  reduced Planck mass. The
coupling of gravity  to other energy and matter  degrees of freedom is
then determined by the effective Planck mass $F(\varphi)$.

Varying the  action Eq.~\eqref{action} with respect to  the metric, we
obtain the Einstein equations, which are given by
\be
\label{Einsteinequation}
  G_{\mu \nu} = \frac{1}{F(\varphi)} \bigg[ \varphi_{,\mu}
      \varphi_{,\nu}
    - \frac{1}{2} g_{\mu \nu} g^{\alpha \beta} \varphi_{,\alpha}
      \varphi_{,\beta}
    + F_{;\mu\nu}(\varphi)
    - g_{\mu \nu} \Box F(\varphi)
    - g_{\mu \nu} U(\varphi) \bigg]\;,
\ee
where $G_{\mu\nu}$ is the Einstein tensor, $\Box$ is the d'Alembertian
operator  and   $,\mu$  and   $;\mu$  denote  partial   and  covariant
derivatives  with  respect  to  the  spacetime  coordinate  $x^{\mu}$,
respectively.   Varying the action  with respect  to the  scalar field
$\varphi$  yields the  Klein-Gordon equation,  written  in Brans-Dicke
form as
\be
\label{Brans-Dicke}
  2 \varpi(\varphi) \Box \varphi(\varphi) =
    - \varpi_{,\varphi}(\varphi) g^{\mu \nu} \varphi_{,\mu}
      \varphi_{,\nu}
    - 4 F_{,\varphi} (\varphi) U(\varphi)
    + 2F(\varphi) U_{,\varphi}(\varphi)\;,
\ee
where  we have  defined  $\varpi(\varphi) =  F(\varphi) +  \frac{3}{2}
F_{,\varphi}^2  (\varphi)$,   in  which  $,\varphi$   denotes  partial
differentiation   with  respect   to  the   scalar   field  $\varphi$.
Hereafter, we will omit arguments on the functions of the scalar field
for notational convenience.

\subsection{Background}

We  shall  assume a  spatially  homogeneous  and isotropic  background
spacetime,  described  by  the Friedmann-Robertson-Walker  (FRW)  line
element
\be
\label{FRW}
  \mathrm{d} s^2 = -\mathrm{d} t^2
    + a^2(t) \delta_{ij} \mathrm{d}x^i \mathrm{d}x^j\;,
\ee
where $\delta_{ij}$  is the  Kronecker delta and  $a(t)$ is  the scale
factor.  In FRW space-time, Eq.~\eqref{Brans-Dicke} is then
\be
\label{KG}
  \ddot{\varphi} + 3H \dot{\varphi} = \frac{1}{2 \varpi}
    [- \varpi,_{\varphi} \dot{\varphi}^2
    + 4 F,_{\varphi} U - 2 F U,_{\varphi} ]\;,
\ee
where $\dot{}$ denotes differentiation with respect to the cosmic time
$t$ and  $H = \dot{a}/a $  is the Hubble  parameter.  Furthermore, the
Friedmann equations take the form
\begin{subequations}
\label{fried}
\begin{align}
  H^2 & = \frac{1}{3F} \bigg[ \frac{1}{2} \dot{\varphi}^2
    + U - 3H \dot{F} \bigg]\;,
  \\
  - 2 \dot{H} & = \frac{1}{F} \bigg[ \dot{\varphi}^2
    + \ddot{F} - H\dot{F}  \bigg]\;.
\end{align}
\end{subequations}
Equations   \eqref{Einsteinequation},  \eqref{KG} and \eqref{fried}
suggest to define an effective  energy density $\rho$ and pressure $p$
for the scalar field as follows:
\begin{subequations}
\begin{align}
  \rho & = \frac{1}{F } \bigg[ \frac{1}{2} \dot{\varphi}^2
    + U - 3H \dot{F} \bigg]\;,
  \\
  p & = \frac{1}{F } \bigg[ \frac{1}{2} \dot{\varphi}^2
    - U + \ddot{F} + 2H \dot{F} \bigg]\;.
\end{align}
\end{subequations}
We note that these are effective quantities and that the corresponding
energy-momentum   tensor    $T^{(\varphi)}_{\mu\nu}$   is   conserved,
i.e.~$T_{\mu\nu}^{(\varphi);\mu} = 0$.

In order  to test  the generalities of  the single--field  results, we
consider a two-field model,  in which the action Eq.~\eqref{action} is
supplemented with an  additional minimally-coupled scalar $\chi$ with
action
\be
  \label{twofieldS}
  S^{(\chi)} \ =\  -\:\frac{1}{2}\int d^4 x \sqrt{-g} \bigg[
  g^{\mu \nu} \chi_{,\mu} \chi_{,\nu} + 2V(\chi)  \bigg],
\ee
where the potential $V(\chi)$ is given by
\be
  V(\chi)\ =\ \frac{1}{2} m_{\chi}^2 \chi^2\;.
\ee
The  pertinent background  field  and perturbation  equations for  the
two-field   model  $S=S^{(\varphi)}+S^{(\chi)}$   are   summarized  in
App.~\ref{sec:2f}.

\subsection{Perturbations}

\subsubsection{Scalar perturbations}

We  will now  focus  our  attention on  the  first-order  perturbation
equations,  which will  be studied  in the  Newtonian gauge.   In this
gauge, the scalar metric  perturbations are expressed by the following
line element, cf.~Eq.~\eqref{FRW},
\be
  d s^2 =  -( 1 + 2 \Psi ) \mathrm{d} t^2
    + a(t)^2(1 - 2 \Phi) \delta_{ij} \mathrm{d}x^i \mathrm{d}x^j\;,
\ee
where $\Psi$ and $\Phi$ are the scalar metric perturbations.

The scalar  field $\varphi(t,\mathbf{x})$ is  decomposed in terms
of  the  homogeneous  background  contribution  $\varphi(t)$  and  the
perturbation $\delta\varphi(t,\mathbf{x})$, i.e.
\be
  \varphi(t, \mathbf{x}) = \varphi (t)
    + \delta \varphi (t, \mathbf{x} )\;.
\ee
Thereafter, we work with  the Fourier components of the perturbations,
$ \delta  \varphi_k(t) $, satisfying  $ \nabla^2 \delta \varphi_k  = -
k^2 \delta \varphi_k  $. In what follows, the subscript $  k $ will be
omitted in order to shorten the subsequent expressions.

The resulting perturbation equation for the scalar field is
\begin{align}
  \delta \ddot{\varphi} & + \bigg[ 3H + \frac{\varpi,_{\varphi}}{\varpi}
      \dot{\varphi} \bigg] \delta \dot{\varphi}
    + \bigg[ \frac{1}{2}
      \bigg( \frac{\varpi,_{\varphi}}{\varpi} \bigg)_{\!\!,\varphi}\!
      \dot{\varphi}^2
    - \frac{1}{2} \bigg( \frac{1}{\varpi} ( 4F,_{\varphi}U
      - 2FU,_{\varphi}  ) \bigg)_{\!\!,\varphi} + \frac{k^2}{a^2} \bigg]
    \delta \varphi
  \nonumber \\
  & - ( \dot{\Psi} + 3 \dot{\Phi} ) \dot{\varphi}
    + \frac{1}{\varpi} \bigg[ -4F_{,\varphi} U  + 2F U_{,\varphi}  \bigg]
    \Psi = 0\; .
\end{align}
Additionally, in the Newtonian gauge, the perturbed Einstein equations
are given by the following:
\begin{subequations}
\label{Einseq}
\begin{align}
  3H( \dot{\Phi} + H \Psi) + \frac{k^2}{a^2} \Phi
    & = - \frac{1}{2} \delta \rho\;,
  \\
  \dot{\Phi} + H \Psi & = - \frac{1}{2} \delta q\;,
  \\
  \ddot{\Phi} + (2 \dot{H} + 3H^2 ) \Psi
    + H( \dot{\Psi} + 3 \dot{\Phi} ) & = \frac{1}{2} \delta p\;,
\end{align}
\end{subequations}
where  $\delta\rho$, $\delta  q$  and $\delta  p$,  obtained from  the
effective energy-momentum  tensor $  T_{\mu \nu} $  mentioned earlier,
are the  perturbations in the  energy density, momentum  potential and
pressure, respectively:
\begin{subequations}
\begin{align}
  \delta \rho & = \frac{1}{F} \bigg[ \dot{\varphi} \delta \dot{\varphi}
    - \dot{\varphi}^2 \Psi
    + U,_\varphi \delta \varphi   	
    + \, 3 \dot{F} ( \dot{\Phi} + 2H \Psi )
    - 3H ( \delta \dot{F} + H \delta F )
    - \frac{k^2}{a^2} \delta F \bigg]\;,
  \\
  \delta q & = - \frac{1}{F} \bigg[\dot{\varphi} \delta \varphi
    + \delta \dot{F} - \dot{F} \Psi - H \delta F \bigg]\;,
  \\
  \delta p & = \frac{1}{F} \bigg[ \dot{\varphi} \delta \dot{\varphi}
    - \dot{\varphi}^2 \Psi
    - U,_\varphi \delta \varphi   		
    - \,  p \, \delta F + \delta \ddot{F} + 2H \delta \dot{F}
    - \dot{F} \dot{\Psi} - 2 \dot{F} \dot{\Phi}	\nonumber 	\\
    & \qquad - \, 2 ( \ddot{F} + 2H \dot{F} ) \Psi
    + \frac{k^2}{a^2} \delta F \bigg]\;.
\end{align}
\end{subequations}
Hence, we find that anisotropic  stress is present in the Jordan frame
with
\be
  \Phi - \Psi = \frac{ \delta F }{F}
    = \frac{F_{,\varphi} \delta \varphi}{F}\;.
\ee

The observational quantities  include the spectral  index $
n_s  $ and its  running $  \alpha $,  which can  be obtained  from the
curvature       (scalar)       power       spectra      by       using
\cite{Kosowsky:1995aa,Hinshaw:2012aka}
\be
\label{spectralindex}
  \mathcal{P}_{\zeta} (k) = \mathcal{P}_{\zeta} (k_0)
    \bigg( \frac{k}{k_0} \bigg)^{n_s(k_0) - \, 1
    + \frac{1}{2} \ln ( k/k_0 ) \, \alpha }\;.
\ee
The scalar perturbations
\be
  \label{scal}
  \mathcal{P}_{\zeta}\ =\ \frac{k^3}{2 \pi^2} | \zeta |^2
\ee
are provided by the curvature perturbation on constant hypersurfaces 
$\zeta $, defined via
\be
  \zeta = - \, \Phi - \frac{H}{\dot{\rho}} \delta \rho\;.
\ee
At the Planck pivot scale $k_0$, the amplitude of the power spectrum is
$   \mathcal{P}_{\zeta}    (k_0)   \sim   2.15    \times   10^{-9}   $
\cite{Ade:2013uln}. The running  index $ \alpha $, in  relation to the
spectral index $ n_s $, is given by
\be
  \alpha = \frac{d \, n_s}{d \, \mathrm{ln} k}\;.
\ee
The  current best  fit  values for  both  the spectral  index and  its
running, as measured by Planck \cite{Ade:2013uln}, are
\be
  n_s = 0.9603 \pm 0.0073~, \qquad \qquad
  \alpha = -\,0.0134 \pm 0.0090\;.
\ee

\subsubsection{Tensor perturbations}

We shall  also study the  effect of variations  in the Planck  mass on
tensor  perturbations.  The equations  for the  tensor modes  take the
standard  form, since they  are not  affected by  the presence  of the
non-minimal coupling  $F(\varphi)$.  Specifically, the  power spectrum
for the tensor perturbations is given by \cite{Habib:2005mh}
\be
\label{tens}
  \mathcal{P}_{\mathrm{T}} \ =\  \frac{k^3}{2 \pi^2}
  \bigg| \frac{v_k}{a} \bigg|^2~,
\ee
where the mode equation for gravitational waves takes the form 
\be
  v_k'' + \bigg( k^2 - \frac{a''}{a} \bigg) v_k\ =\  0\;,
\ee
in which the prime ($'$) denotes the derivative with respect to conformal time $\eta=\int_0^t\frac{\mathrm{d}t'}{a(t')}$. 

Finally, the  tensor-to-scalar ratio $  r $ is  given in terms  of the
scalar  and   tensor  power  spectra   $\mathcal{P}_{\mathrm{T}}$  and
$\mathcal{P}_{\zeta}$,  see  Eqs.~\eqref{scal}  and  \eqref{tens} via
\cite{PeterUzan}
\be
  r = \frac{8\mathcal{P}_{\mathrm{T}} }{ \mathcal{P}_{\zeta} }~.
\ee

\section{Model with step variation in the Planck mass}
\label{sec:model}

In this section, we will consider models in which the effective Planck
mass  undergoes a step  transition during  the inflationary  epoch. To
this end, we consider the following non-minimal coupling and potential
for a canonical Brans-Dicke scalar field:
\begin{align}
\label{eq:Fdef}
  F(\varphi) & = 1-\beta \,
    \{\,  1 + \tanh[(\varphi - \varphi_\ast)/\gamma\,]\,\}\;,
  \\
  \label{eq:Udef}
  U(\varphi) & = \frac{1}{2} m_{\varphi}^2 \varphi^2\;,
\end{align}
where $ m_{\varphi} $  is the mass of the scalar field  $ \varphi $, $
\beta$ is  a dimensionless constant, and $\gamma$  and $\varphi_\ast $
are  constants of  mass dimension.  As  we shall  see, the  parameters
$\beta$  and $\gamma$  determine the  amplitude and  sharpness  of the
transition  in $F$  and $\varphi_\ast$  determines the  field value at
which  the  transition  occurs. We have chosen the quadratic potential for concreteness. However, we should emphasize that the features observed in the forthcoming sections are anticipated to persist for other choices of the potential $U(\varphi)$. 

In the  first instance, we  will consider a single field
model,    in    which    the    Brans-Dicke   field    also    drives
inflation. Subsequently, we will  consider a two-field model, in which
a  second minimally-coupled scalar field acts  as the  inflaton. Nevertheless, in both cases and for  each of  the  benchmark  models
considered, the  values of the parameters  are chosen so  as to obtain
successful inflation, with the  inflationary period lasting a total of
66 $e$-folds.   

Before proceeding, we will now illustrate that the Jordan-frame model described above is \emph{not} equivalent to an Einstein-frame model of an inflaton potential with a step, see \cite{Adams:2001vc}. By  means of  a conformal  transformation, we  could transform the model in Eqs.~\eqref{eq:Fdef} and \eqref{eq:Udef} to the
Einstein frame.  Therein, the new potential for  the Brans-Dicke field
$\varphi$ would become
\be\label{Utilde}
  \tilde{U}(\varphi)\ =\ \frac{U(\varphi)}{F^2(\varphi)}\ =\
  \frac{m_{\varphi}^2\varphi^2}{2\big\{1-\beta \,
    \{\,  1 + \tanh[(\varphi - \varphi_\ast)/\gamma\,]\}^2}\;,
\ee
which  resembles   the  step  potential   in  \cite{Adams:2001vc}  for
$\beta\ll  1$.   Note  however  that we  should  anticipate  different
dynamics, since in the Einstein frame the kinetic term of the model we
consider   will   not   be   of   canonical  form,   as   it   is   in
\cite{Adams:2001vc}.  The canonical field $\tilde{\varphi}$ is related
to the non-canonical $\varphi$ through the relation
\be
  \tilde{\varphi}=\int
  \sqrt{\frac{2F(\varphi)+3
  {F_{,\varphi}(\varphi)}^2}{2F(\varphi)^2}}~d\varphi.
\ee
Inverting the  above equation,  one can derive  $\varphi$ in  terms of
$\tilde{\varphi}$,  at  least  implicitly.   Upon replacement  of
$\varphi$ in  terms of $\tilde{\varphi}$  in $\tilde{U}(\varphi)$, the
potential   for  the  canonical   field  $\tilde{\varphi}$   could  be
obtained. However, the final form of the potential, written in terms of the canonical field $\tilde{\varphi}$, will not be the potential with a step Eq.~\eqref{Utilde}. As such, we conclude that a step potential of a canonical field is not an appropriate phenomenological model of a Jordan-frame action in which the effective Planck mass undergoes a step change. Thus, the model under investigation here differs from those considered previously in the literature, leading to significantly different predictions for the scalar and tensor spectra.

Returning to the Jordan frame, the dynamics of  the fields will be solved numerically, following the
method  outlined  in \cite{Tsujikawa:2002qx};  the  derivatives of  the
background fields  are given their  slow-roll values and  the initial
field  perturbations will have  the standard  oscillatory Bunch-Davies
initial  conditions \cite{Adams:2001vc}.   In order  to  calculate the
tensor perturbations  generated by the  system, we employ  the methods
described in \cite{Adams:2001vc, Habib:2005mh}.

In App.~\ref{sec:app1}, we use an approximate analytic solution to
the background  field equations in order to  illustrate the dependence
on the  parameters $\beta$ and  $\gamma$ of the resulting  features in
the slow-roll  parameter $\varepsilon$. The latter allows  us to infer
the dependency  on the same parameters  of the features  in the scalar
power  spectra $\mathcal{P}_{\zeta}$  and  the tensor-to-scalar  ratio
$r$.

\subsection{Single-field model}

\subsubsection{Minimally-coupled limit}

\begin{figure}[t]
  \begin{center}
    \scalebox{0.85}{\includegraphics{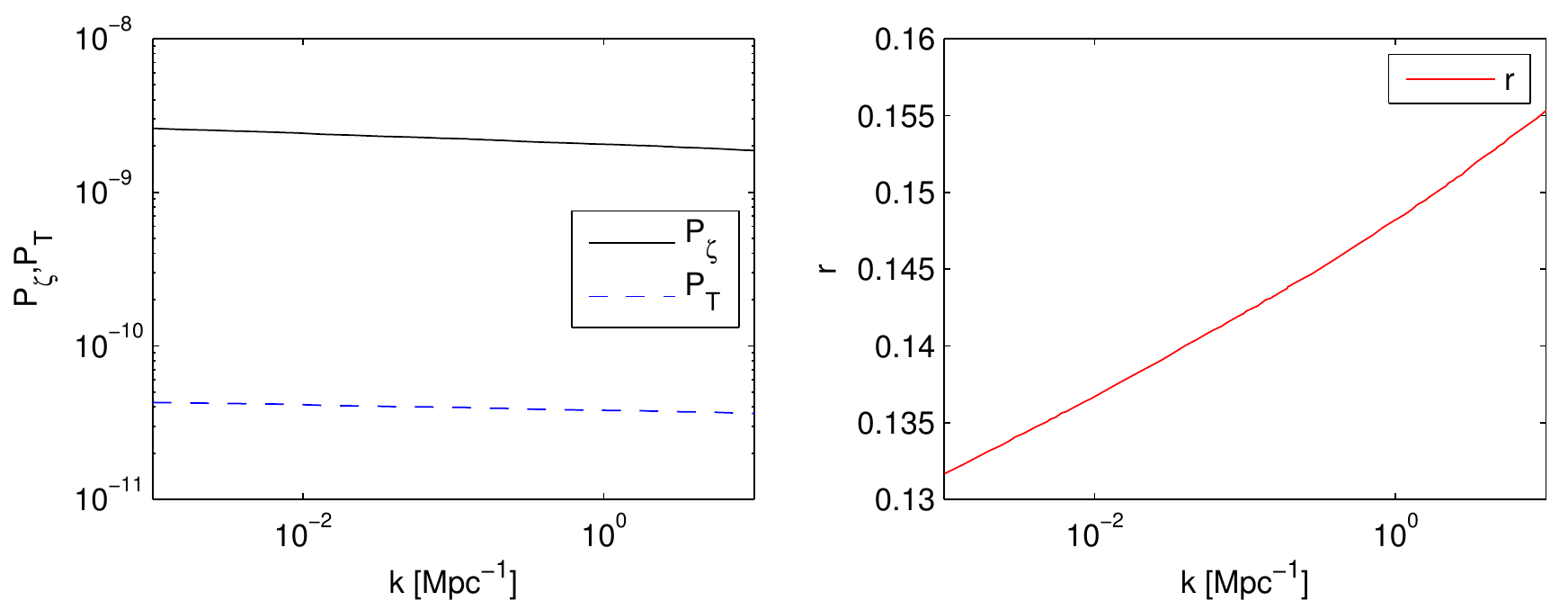}}
    \caption{Power      spectra      (\emph{left})      of      scalar
    ($\mathcal{P}_{\zeta}$,  solid  black)  and  tensor  perturbations
    ($\mathcal{P}_{\mathrm{T}}$,  blue   dashed)  and  the  associated
    tensor-to-scalar ratio  $r$ (\emph{right}) against  wavenumber $k$
    (${\rm Mpc^{-1}}$) for the  minimally-coupled scenario.  The model
    parameters  are  $  m_\varphi  =  6.5 \times  10^{-6}  $,  with  $
    \varphi_{\mathrm{ini}} = 16.179 $. \label{minpowr}}
  \end{center}
\end{figure}

We first  consider the minimally-coupled  case, in which $\beta  = 0$,
i.e.~$ F = 1 $. The value of the scalar field $ \varphi $ at the start
of inflation is  taken to be $ \varphi_{\mathrm{ini}}  = 16.179 $. The
mass of  the scalar field is chosen  to be $ m_{\varphi}  = 6.5 \times
10^{-6} $, so that the  power spectrum for the scalar perturbations at
the Planck  pivot scale  is approximately $  2.15 \times 10^{-9}  $.  The
power spectra  for the  scalar and tensor  perturbations are  given in
Fig.~\ref{minpowr}.   Notice  that  we  have  defined  the  number  of
$e$-folds $N$ such that $N=0$ at the \emph{start} of inflation.

There are no features generated in  this model, as we would expect for
minimally-coupled single-field  inflation. The  spectral  and running
indices are calculated to be
\be
  n_{s} = 0.968865\;, \qquad \alpha = 0.00107427\;.
\ee
In addition to this, the tensor-to-scalar ratio at the  pivot scale $k_0$ is
\be
  r(k_0) = 0.133205
\ee
We  have chosen  a quadratic  potential for  simplicity. The  model is
under  slight pressure from  the Planck  experiment \cite{Ade:2013uln}
(cf.~\cite{Ashoorioon:2013eia}, which attempts to reconcile  this model
with the Planck data), although it  is still within the $68\%$ C.L. in
the $n_s-r$  plane. Nevertheless, the phenomenological conclusions presented later in this paper do not depend heavily on the choice of potential.

\subsubsection{Benchmark 1}

\begin{figure}[t]
  \begin{center}
    \scalebox{0.85}{\includegraphics{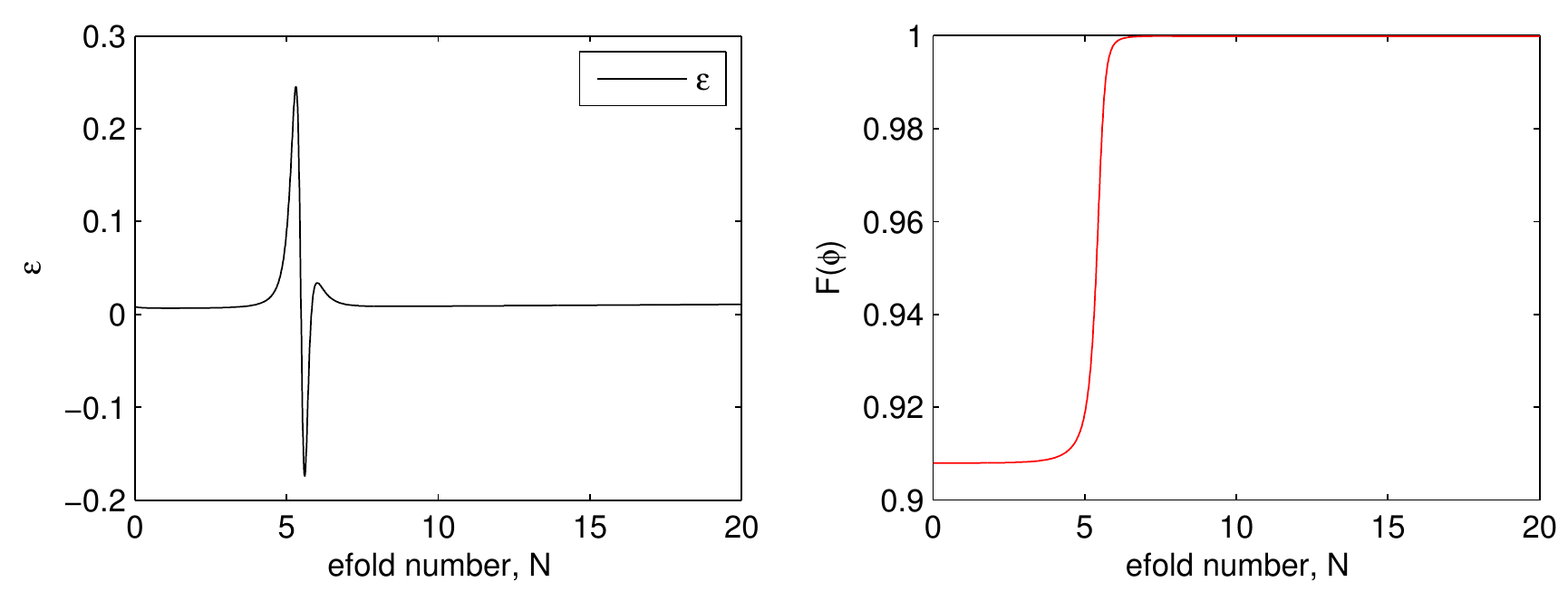}}
    \caption{Evolution  of  the slow-roll  parameter  $ \varepsilon  $
    (\emph{left}) and  the effective  Planck mass $  F$ (\emph{right})
    for the first $ 20 $  $e$-folds of inflation in benchmark model 1.
    The  model parameters  are $  m_\varphi  = 2.1  \times 10^{-5}  $,
    $\beta=0.0460$, $\gamma=0.145$ and $ \varphi_\ast = 15.8 $, with $
    \varphi_{\mathrm{ini}} = 16.5783 $.\label{m1bkgr}}
  \end{center}
\end{figure}

\begin{figure}[t]
  \begin{center}
    \scalebox{0.85}{\includegraphics{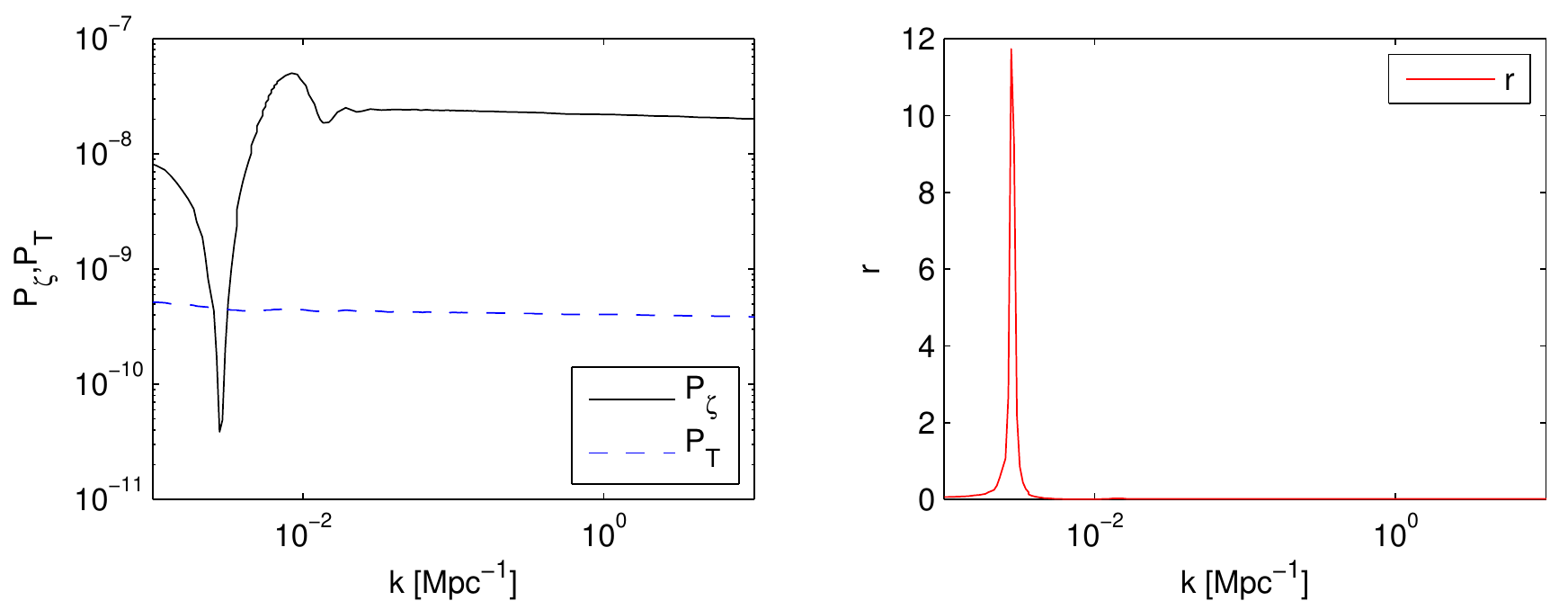}}
    \caption{Power      spectra      (\emph{left})      of      scalar
    ($\mathcal{P}_{\zeta}$,  solid  black)  and  tensor  perturbations
    ($\mathcal{P}_{\mathrm{T}}$,  blue   dashed)  and  the  associated
    tensor-to-scalar ratio  $r$ (\emph{right}) against  wavenumber $k$
    (${\rm Mpc^{-1}}$)  for benchmark  model 1.  The  model parameters
    are  $   m_\varphi  =   2.1  \times  10^{-5}   $,  $\beta=0.0460$,
    $\gamma=0.145$   and   $   \varphi_\ast   =   15.8   $,   with   $
    \varphi_{\mathrm{ini}} = 16.5783 $.\label{m1powr}}
  \end{center}
\end{figure}

\begin{figure}[t]
  \begin{center}
    \scalebox{0.45}{\includegraphics{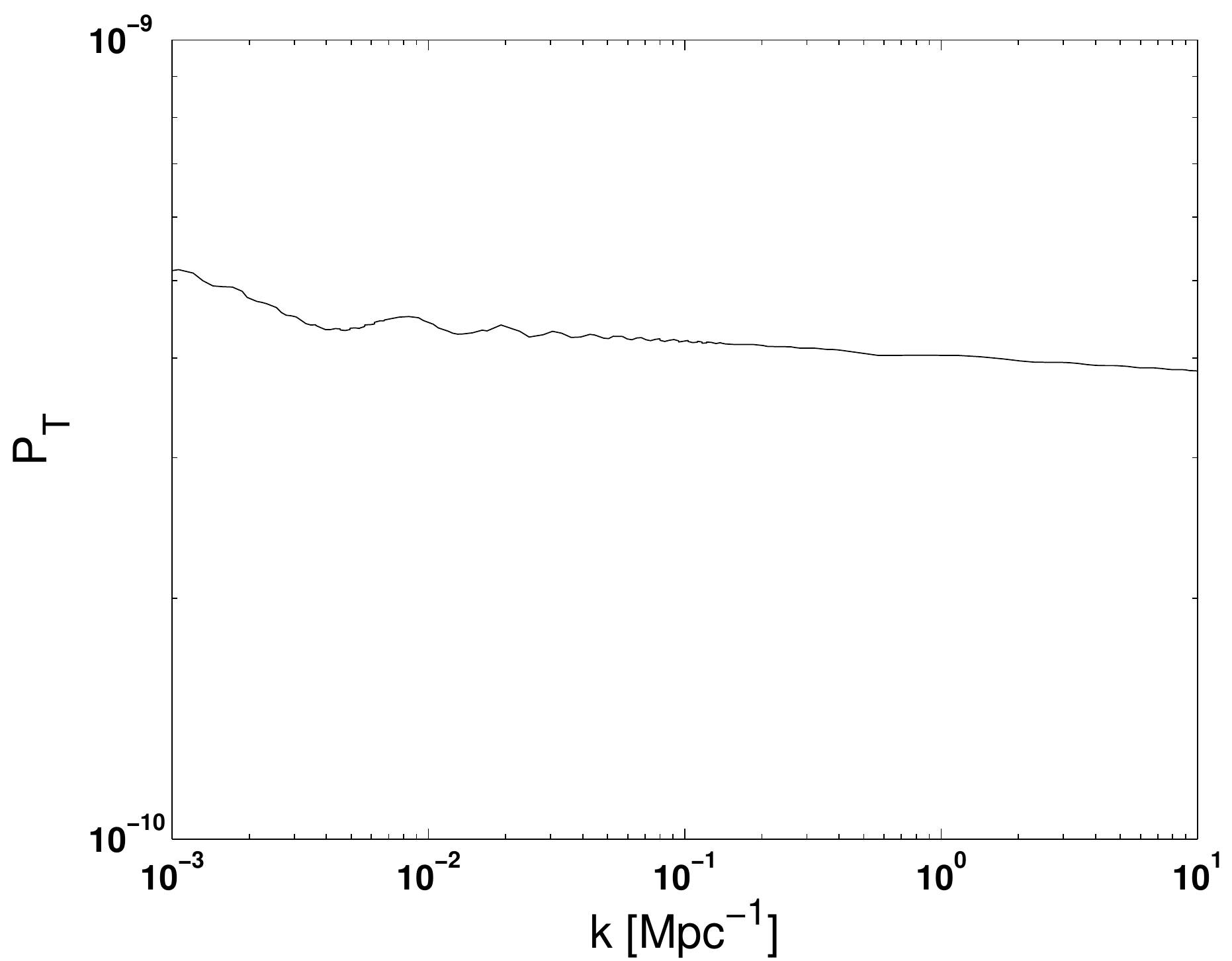}}
    \caption{The   power   spectrum   of   tensor   perturbations   ($
    \mathcal{P}_{\mathrm{T}}   $)  showing  the   damped  oscillations
    present  for  benchmark  model  1.   The model  parameters  are  $
    m_\varphi =  2.1 \times 10^{-5}  $, $\beta=0.0460$, $\gamma=0.145$
    and  $ \varphi_\ast  =  15.8 $,  with  $ \varphi_{\mathrm{ini}}  =
    16.5783 $.\label{m1powtens}}
  \end{center}
\end{figure}

In this first benchmark model,  we will consider a steep transition in
the Planck mass, causing a  violent feature in the slow-roll parameter
$ \varepsilon $. Specifically, the  model parameters are $ m_\varphi =
2.1  \times  10^{-5} $,  $  \beta =  0.0460  $,  $\gamma=0.145$ and  $
\varphi_\ast   =   15.8   $,   with   the  initial   field   value   $
\varphi_{\mathrm{ini}} = 16.5783 $. The $e$-fold evolutions of $ F$ and
$ \varepsilon $ are displayed in Fig.~\ref{m1bkgr}, in which we see a transient violation of slow-roll.\footnote{For a discussion of the calculation of correlation functions for single-field models that does not rely on slow-roll approximations, see for instance \cite{Ribeiro:2012ar}.}

The resulting  scalar and  tensor power spectra  are displayed  in the
left panel of  Fig.~\ref{m1powr}.  We see an extremely  sharp dip in
the power  spectrum of the  scalar perturbation at  $ k \sim  0.003~\mathrm{Mpc}^{-1} $. This feature and the smaller oscillation-like fluctuations that follow coincide with those observed in the slow-roll parameter in Fig.~\ref{m1bkgr}. Notice however that these features do not resemble the dramatic oscillations seen in inflationary models with a step potential, see \cite{Adams:2001vc}.  For this  set  of parameters,  we also observe damped oscillations in the tensor  power spectrum, as  demonstrated in
Fig.~\ref{m1powtens}. However, the amplitude of this effect is significantly smaller than that of the feature in the scalar power spectrum.

These observations may be understood in terms of the behaviour of the slow-roll parameter $\varepsilon$. Specifically, in the single-field model, the scalar power spectrum $\mathcal{P}_{\zeta}\sim H^2/\varepsilon$, whereas $\mathcal{P}_{T}\sim H^2$. Hence, we see that the sharp rise in the slow-roll parameter leads to a sharp dip in the scalar power spectrum at the same scale, whilst leaving the tensor power spectrum largely unaffected.

The tensor-to-scalar ratio as a function  of the wavenumber is presented in the
right-hand side  panel of  Fig.~\ref{m1powr}. We see a sharp rise in $r$ at scales corresponding to the feature in the slow-roll parameter $\varepsilon$, since $r\sim \mathcal{P}_{T}/\mathcal{P}_{\zeta}\sim\varepsilon$.  Although an
unrealistically-large tensor-to-scalar  ratio is generated in the  region of the
Planck pivot scale in this benchmark model, this can  be useful in constraining the parameters of the non-minimal coupling $F$.

\subsubsection{Benchmark 2}

In this  example, we show that one can produce features in the scalar power spectrum that reduce the
tension between the tensor-to-scalar ratio $r$ observed by the Planck
\cite{Ade:2013uln} and  BICEP2 \cite{Ade:2014xna} experiments, as described in Sec.~\ref{intro}. To  this  end,  we  choose  the  following  set  of  model
parameters: $  m_{\varphi} =  6.9 \times 10^{-6}  $, $\beta  = 0.002$,
$\gamma=0.111$ and  $ \varphi_\ast =  15.49 $, with the  initial field
value $ \varphi_{\mathrm{ini}} = 16.2271 $.

\begin{figure}[t]
  \begin{center}
    \scalebox{0.75}{\includegraphics{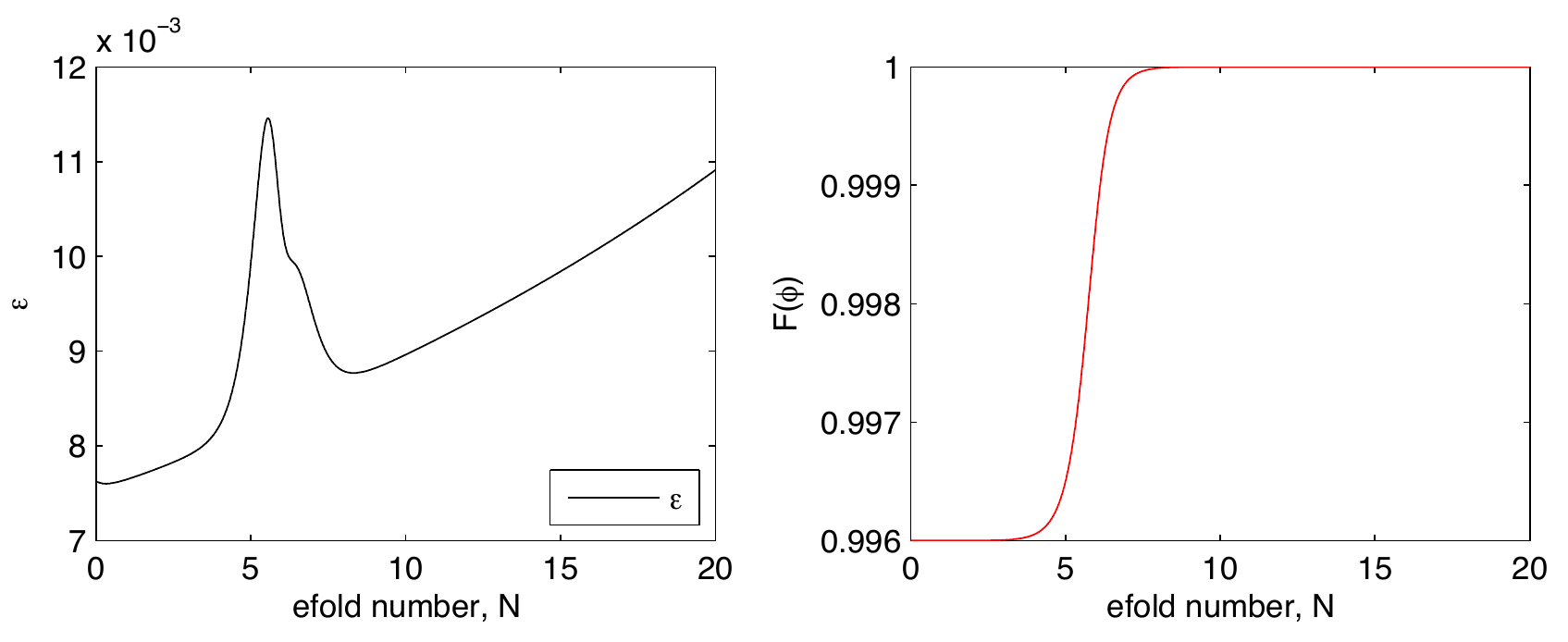}}
    \caption{Evolution  of  the slow-roll  parameter  $ \varepsilon  $
    (\emph{left}) and  the effective  Planck mass $  F$ (\emph{right})
    for the first $ 20 $  $e$-folds of inflation in benchmark model 2.
    The  model parameters  are $  m_\varphi  = 6.9  \times 10^{-6}  $,
    $\beta=0.002$, $\gamma=0.111$ and $ \varphi_\ast = 15.49 $, with $
    \varphi_{\mathrm{ini}} = 16.2271 $.\label{m2bkgr}}
  \end{center}
\end{figure}

In Fig.~\ref{m2bkgr}, we see  that the slow-roll parameter creates a
peak due  to the  increase in the  coupling at  approximately $ N  = 6
$. Notice that, with this combination of parameters, the initial value
of $F(\varphi_{\mathrm{ini}}) = 0.996  $ deviates by less than $0.5\%$
from minimal coupling, compared with $\sim10\%$ deviation in benchmark
1.

\begin{figure}[t]
  \begin{center}
    \scalebox{0.75}{\includegraphics{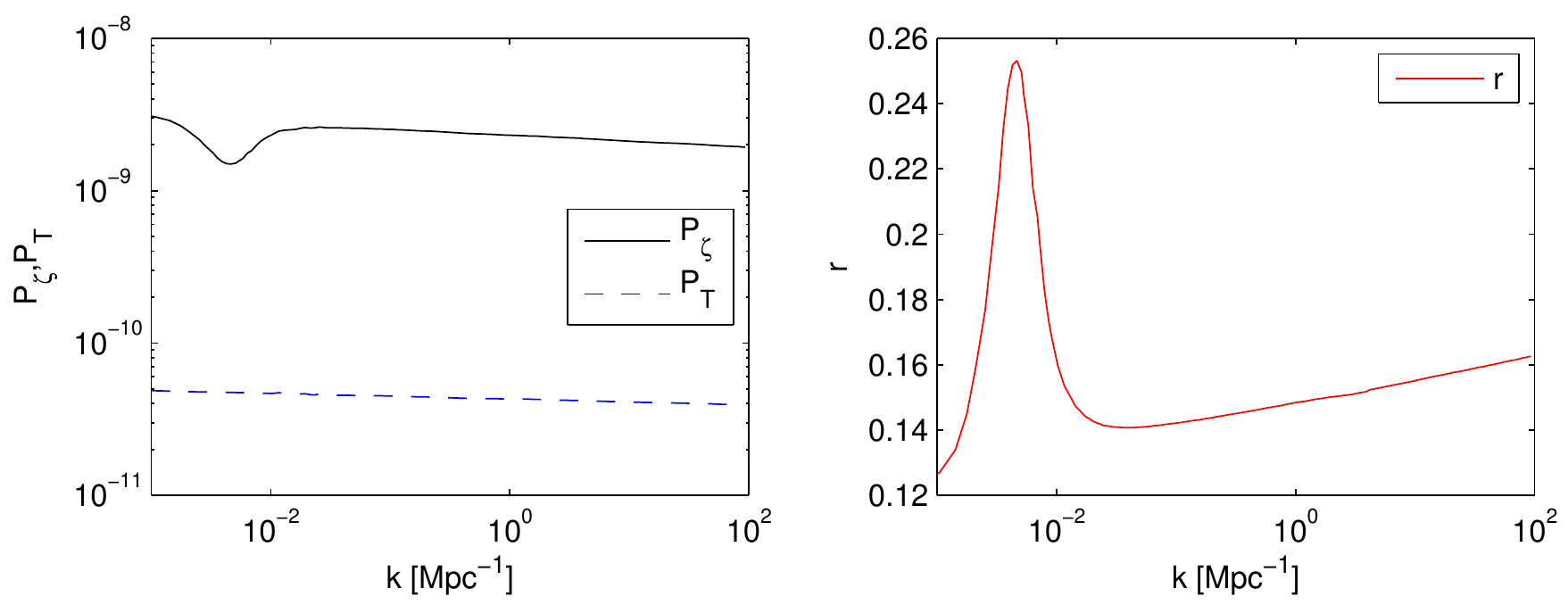}}
    \caption{Power      spectra      (\emph{left})      of      scalar
    ($\mathcal{P}_{\zeta}$,  solid  black)  and  tensor  perturbations
    ($\mathcal{P}_{\mathrm{T}}$,  blue   dashed)  and  the  associated
    tensor-to-scalar ratio  $r$ (\emph{right}) against  wavenumber $k$
    (${\rm Mpc^{-1}}$)  for benchmark  model 2.  The  model parameters
    are   $  m_\varphi   =  6.9   \times  10^{-6}   $,  $\beta=0.002$,
    $\gamma=0.111$   and   $   \varphi_\ast   =  15.49   $,   with   $
    \varphi_{\mathrm{ini}} = 16.2271 $.\label{m2powr}}
  \end{center}
\end{figure}

The resulting  power spectra for  the scalar and  tensor perturbations
are  displayed  in the  left  panel  of  Fig.~\ref{m2powr} and  the
tensor-to-scalar  ratio  versus wavenumber  in  the  right panel.   As
expected,  a reduction  in the  power is  observed.  The  tensor power
spectrum,  on the other  hand, is  unaffected. We  see that,  with this
choice of  parameters, the value  of $r$ at  $k\simeq 0.005~{\rm Mpc}^{-1}$,
see Eq.~\eqref{bicep2r}, is consistent  with the BICEP2 result, whilst
maintaining  agreement  with  the   Planck  limit  of  $r\leq0.11$  at
$k\simeq 0.002~{\rm Mpc}^{-1}$.

For  $k>10^{-2}$, the  value  of  the spectral  tilt  is $n_s  \approx
0.98$. The maximum value of the  spectral index in the vicinity of the
feature is obtained by means of \cite{Stewart:1993bc}
\be
  n_s(k)\ \approx\ 1+k\,
  \frac{\mathrm{d}}{\mathrm{d} k}\ln\,\mathcal{P}_{\zeta}(k)\;.
\ee
One  can obtain  a rough  estimate on  the maximum  magnitude  for the
non-linearity parameter  $f_\mathrm{{NL}}$ in the  squeezed limit \cite{Maldacena:2002vr} of
\be
  |f_{\mathrm{NL}}|_{\mathrm{max}}\ \approx \ \frac{5}{12}|1-n_s|
  \ \approx \ 0.45\;.
\ee
Thus,  even in  the vicinity  of  the transition,  this model  remains
consistent  with  the  Planck   limit  for  local  non-Gaussianity  of
$f_{\mathrm{NL}}^{\mathrm{loc}}\:=\:2.7\pm5.8$ \cite{Ade:2013ydc}.

\subsubsection{Benchmark 3}

\begin{figure}[t]
  \begin{center}
    \scalebox{0.85}{\includegraphics{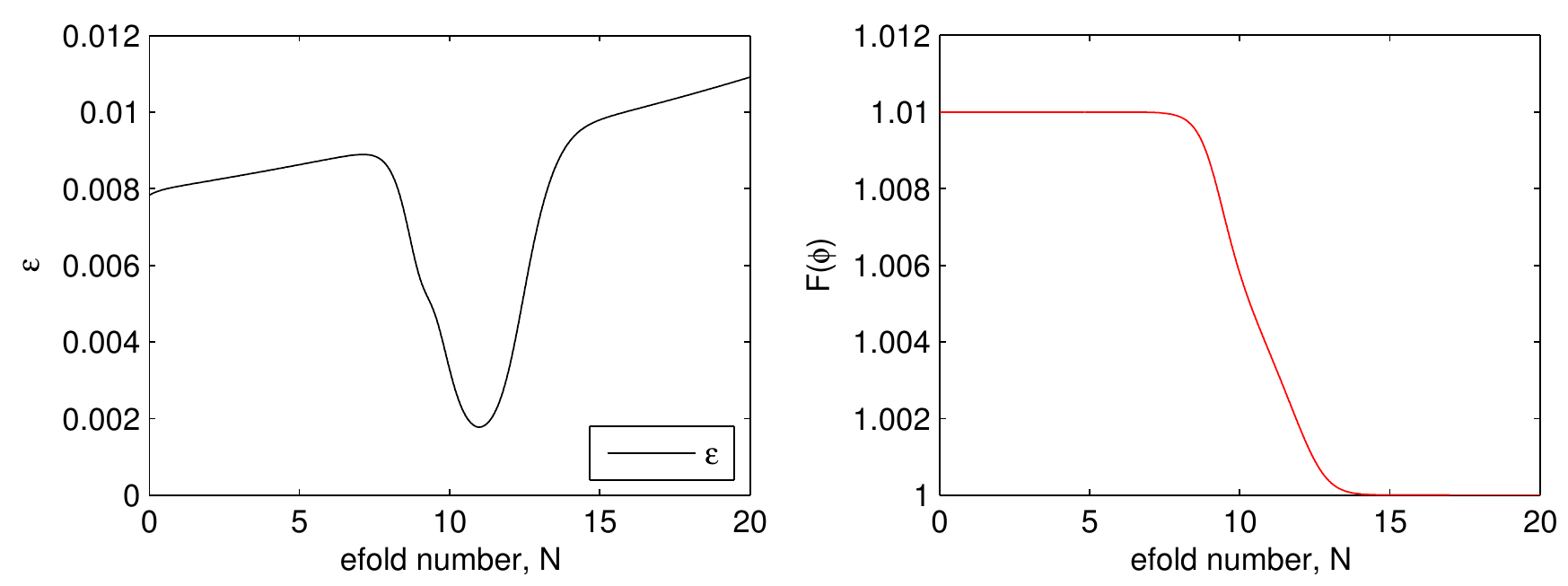}}
    \caption{Evolution  of  the slow-roll  parameter  $ \varepsilon  $
    (\emph{left}) and  the effective  Planck mass $  F$ (\emph{right})
    for the  first $ 20 $  $e$-folds of inflation  in benchmark model 3. 
    The model  parameters are $
    m_\varphi =  6.5 \times 10^{-6}  $, $\beta=-0.005$, $\gamma=0.100$
    and  $ \varphi_\ast  = 14.64  $, with  $  \varphi_{\mathrm{ini}} =
    15.9055 $.\label{m3bkgr}}
  \end{center}
\end{figure}

\begin{figure}[t]
  \begin{center}
    \scalebox{0.85}{\includegraphics{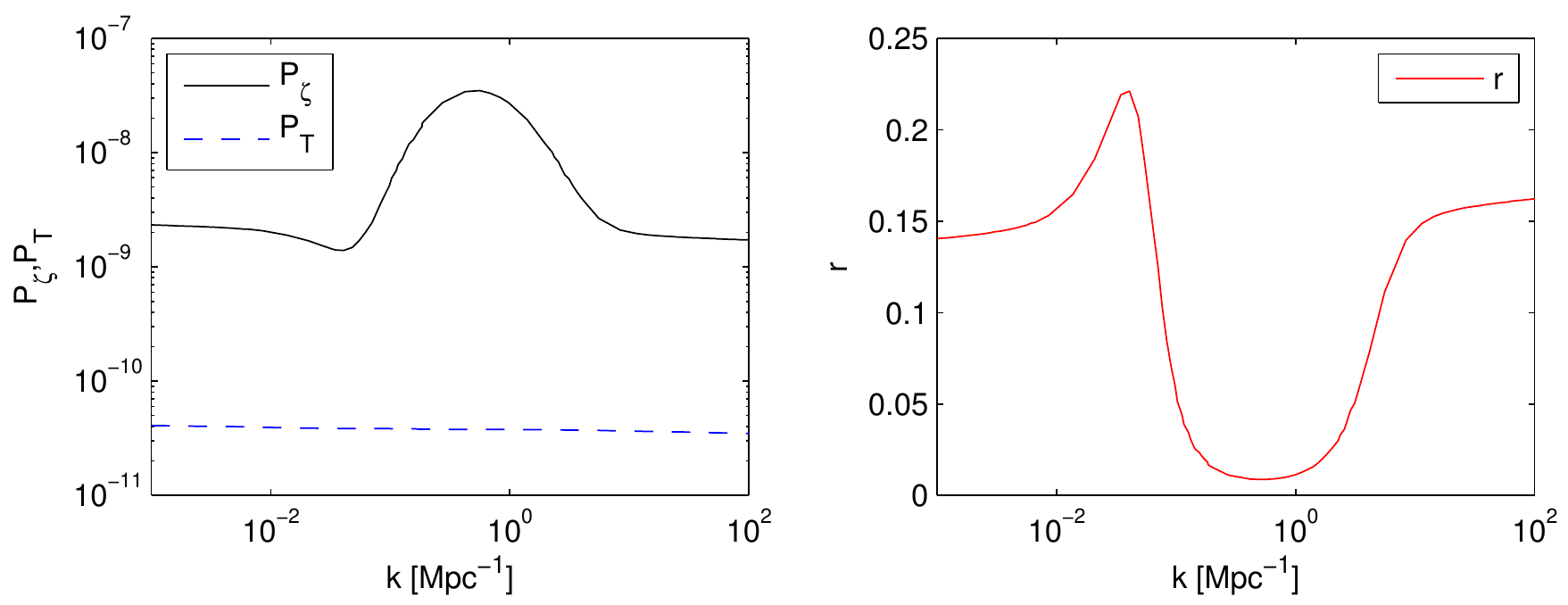}}
    \caption{Power      spectra      (\emph{left})      of      scalar
    ($\mathcal{P}_{\zeta}$,  solid  black)  and  tensor  perturbations
    ($\mathcal{P}_{\mathrm{T}}$,  blue   dashed)  and  the  associated
    tensor-to-scalar ratio  $r$ (\emph{right}) against  wavenumber $k$
    (${\rm Mpc^{-1}}$)  for benchmark  model 3.  The  model parameters
    are  $   m_\varphi  =   6.5  \times  10^{-6}   $,  $\beta=-0.005$,
    $\gamma=0.100$   and   $   \varphi_\ast   =  14.64   $,   with   $
    \varphi_{\mathrm{ini}} = 15.9055 $.\label{m3powr}}
  \end{center}
\end{figure}

Finally, to  illustrate the enhancement of power  over certain scales,
we  consider  the  case  in  which   the  the  factor  $  \beta  $  is
negative. The model  parameters chosen are $ m_{\varphi}  = 6.5 \times
10^{-6} $,  $\beta=-0.005$, $\gamma=0.100$ and $  \varphi_\ast = 14.64
$, with the initial field value $ \varphi_{\mathrm{ini}} = 15.9055 $.

The evolutions of the slow-roll  parameter $ \varepsilon $ and coupling
$  F(\varphi)  $  for  this  choice of  parameters  are  displayed  in
Fig.~\ref{m3bkgr}.   We see that  the change  in sign  of $  \beta $
causes a dip in the  evolution of the slow-roll parameter, which lasts
approximately $7$ $e$-folds  for this set of parameters.   As we would
anticipate given  the earlier  examples, this results  in a  region of
enhancement in the scalar power spectrum. The scalar power spectrum is
shown  in  the  left   panel  of  Fig.~\ref{m3powr}.   In  addition,
Fig.~\ref{m3powr} presents both the  scalar and tensor power spectra
as  well as  the gravitational  coupling as  functions  of wavenumber.
There is no considerable effect  upon the tensor power spectrum.  From
the right panel of Fig.~\ref{m3powr}, we see that the enhancement in
the scalar power spectrum suppresses the tensor-to-scalar ratio in the
range: $ 10^{-1}~{\rm Mpc^{-1}} < k <  10^1~{\rm Mpc^{-1}}$ compared
with the enhancement for $  10^{-2}~{\rm Mpc^{-1}} <k < 10^{-1}~{\rm
Mpc^{-1}} $, corresponding to the initial suppression of power.

\subsection{Two-field model}

In this section, we calculate the power spectra of a two-field model 
directly in the Jordan frame. The transition
in the Planck mass results again from the Brans-Dicke scalar $\varphi$, see
Eq.~\eqref{action}, whilst  inflation is instead driven dominantly by  an additional
minimally-coupled scalar $\chi$, see Eq.~\eqref{twofieldS}.

\begin{figure}[t]
  \begin{center}
    \scalebox{0.85}{\includegraphics{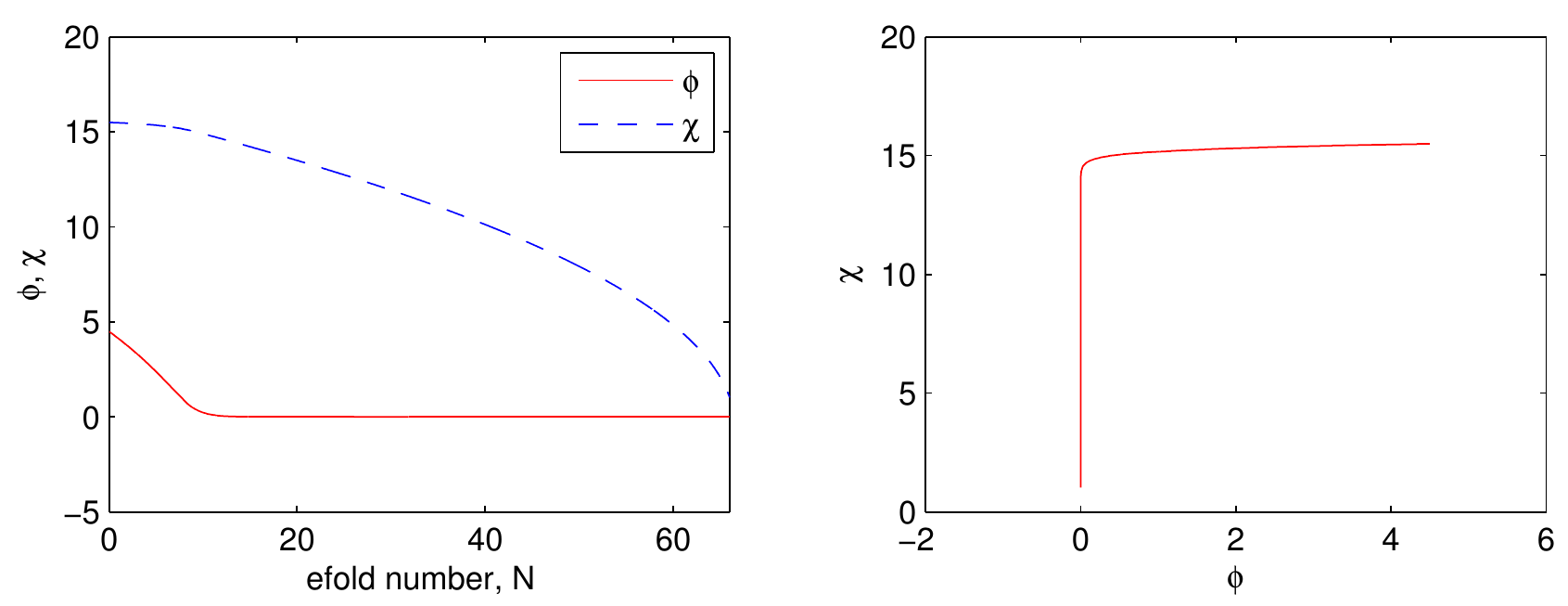}}
    \caption{Evolution of $\varphi$ and $\chi$ versus $e$-fold
    number $N$ (\emph{left}) and the field trajectory  in the $\varphi$-$\chi$ plane for the same $e$-foldings, illustrating the sharp turn at $N\sim 10$ (\emph{right}).\label{twofevo}}
  \end{center}
\end{figure}

\begin{figure}[t]
  \begin{center}
    \scalebox{0.85}{\includegraphics{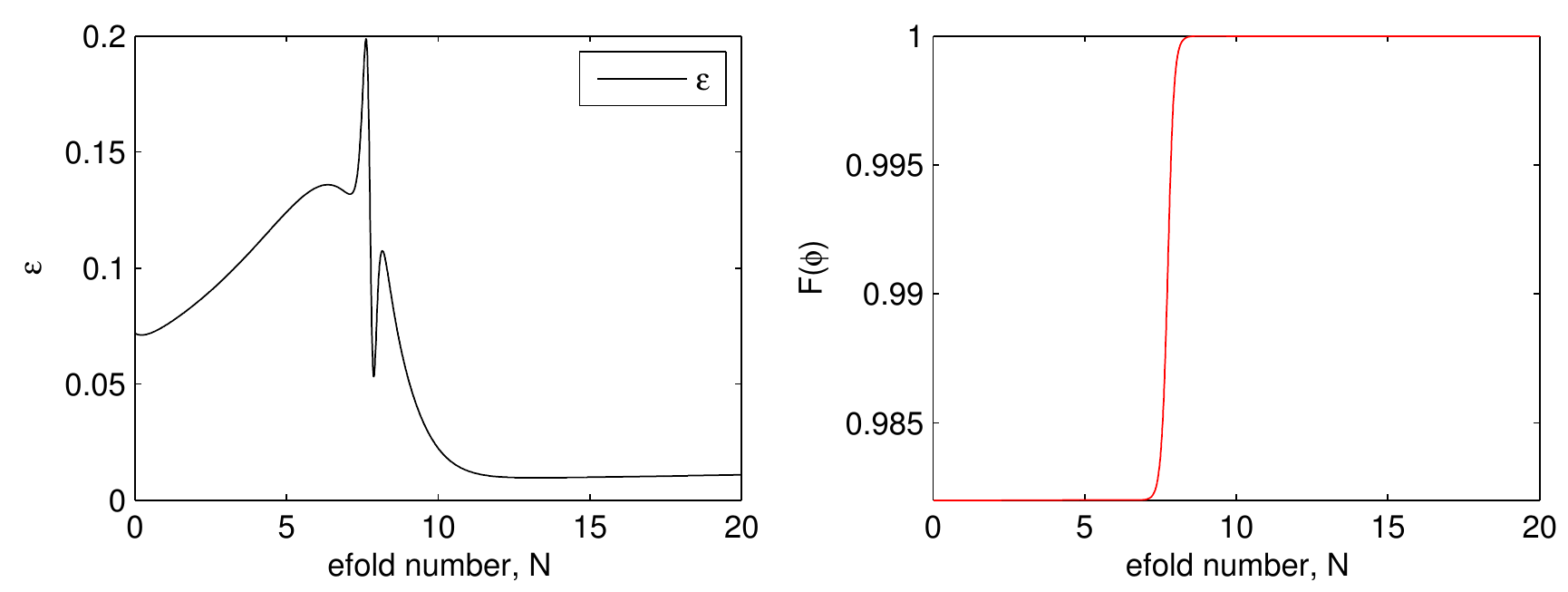}}
    \caption{Evolution  of  the slow-roll  parameter  $ \varepsilon  $
    (\emph{left}) and  the effective  Planck mass $  F$ (\emph{right})
    for  the first  $ 20  $ $e$-folds  of inflation  in  the two-field
    model.  The  model parameters are  $m_{\chi}=5.6\times 10^{-6}$, $
    m_\varphi =  4.48 \times 10^{-5}  $, $\beta=0.009$, $\gamma=0.111$
    and $ \varphi_\ast = 1.00  $, with $ \chi_{\mathrm{ini}} = 15.489$
    and $\varphi_{\mathrm{ini}}=4.5$. \label{2fbkgr}}
  \end{center}
\end{figure}

The   model  parameters   chosen  are   $m_{\chi}=5.6\times  10^{-6}$,
$m_{\varphi}=4.48\times  10^{-5}$,  $\beta=0.009$, $\gamma=0.111$  and
$\varphi_{\ast}=1.0$.      The     initial     field    values     are
$\varphi_{\mathrm{ini}}=4.5$  and $\chi_{\mathrm{ini}}=15.489$. Figure
\ref{twofevo} shows  the evolution of the  Brans-Dicke field $\varphi$
and the scalar $\chi$.  Therein  and for these model parameters, it is
clear  that the last  $50$ $e$-folds  of inflation  are driven  by the
would-be  inflaton  field  $\chi$.   The evolutions  of  the  slow-roll
parameter $\varepsilon$ and the effective Planck mass $F$ are shown in
Fig.~\ref{2fbkgr}. Here,  we see the  feature arising from  the step
change  in the  effective Planck  mass superposed  upon  an additional
background from  the more-rapid evolution of the  Brans-Dicke field, as it rolls to the origin.

The  smooth  enhancement in  the  tensor  power
spectrum for $k\lesssim  10^{-2}~\mathrm{Mpc}^{-1}$ from the change of
slope in the  tensor spectral index can be understood  in terms of the
overall  reduction   of  the  slow-roll  parameter   after  the  turn in the field trajectories shown in Fig.~\ref{twofevo}. Specifically, the  $k$-dependent tilt of the tensor  power spectrum is
given in the Jordan frame by \cite{Hwang:1996xh, Hwang:2001pu, DeFelice:2010aj}
\be
 n_t\ =\ -2\varepsilon-\frac{\dot{\varphi}F_{,\varphi}}{H F}\;,
\ee
where   $\varepsilon=-\dot{H}/H^2$   is   the   usual   first   slow-roll   parameter.  Noting that  the variation  of $F(\varphi)$  is zero
before and after  the transition, the slope of  the tensor spectrum is
given  only by $\varepsilon$, which is  larger before the  turn in the
trajectory  occurs.

\begin{figure}[t]
  \begin{center}
    \scalebox{0.85}{\includegraphics{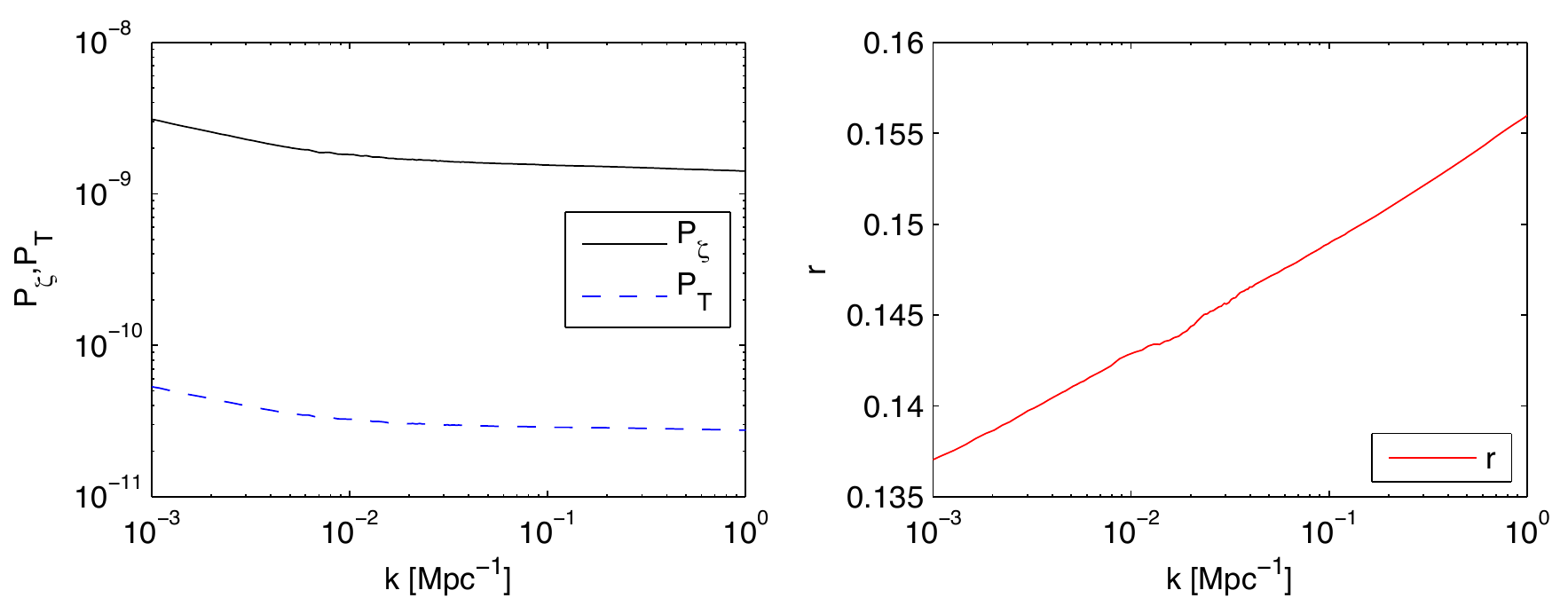}}
    \caption{Power      spectra      (\emph{left})      of      scalar
    ($\mathcal{P}_{\zeta}$,  solid  black)  and  tensor  perturbations
    ($\mathcal{P}_{\mathrm{T}}$,  blue   dashed)  and  the  associated
    tensor-to-scalar ratio  $r$ (\emph{right}) against  wavenumber $k$
    (${\rm Mpc^{-1}}$) for the  two-field model.  The model parameters
    are  $m_{\chi}=5.6\times  10^{-6}$,  $  m_\varphi  =  4.48  \times
    10^{-5} $, $\beta=0.009$, $\gamma=0.111$ and $ \varphi_\ast = 1.00
    $, with  $\chi_{\mathrm{ini}}=4.5$ and $  \varphi_{\mathrm{ini}} =
    15.489 $.\label{2fpowr}}
  \end{center}
\end{figure}

However, as we see from Fig.~\ref{2fpowr}, the variation
in the scalar power  spectra that resulted from the  transition in the Planck
mass in the single-field cases is not apparent. This is in spite of the fact that the fluctuation
of  the slow-roll  parameter  in  the vicinity  of  the transition  is
comparable with  the first  single-field benchmark model. This observation can be understood as follows. The  turn in the field trajectory also  leads  to   the  conversion  of
isocurvature to curvature perturbations, which washes out the anticipated feature and results in the
enhancement  of the  scalar power  spectrum at scales $k\lesssim
10^{-2}~\mathrm{Mpc}^{-1}$, leaving the trajectory  before the turn
\cite{Gordon:2000hv}. This conversion of isocurvature modes for non-minimally-coupled two-field modes renders the curvature perturbations frame-dependent~\cite{White:2012ya}. We have examined this observed suppression of the features in the power spectrum with larger transitions in Newton's constant. In this case, although still partially washed out, the features could be more clearly seen. Further study of such effects on the scalar power spectrum, and also the curvature spectrum in the Einstein frame, will be presented in future work.

\newpage
\section{Conclusions}
\label{sec:con}

We  have studied  the impact  of sudden  transitions in  the effective
Planck mass during inflation on  primordial power spectra. In the case
of single field  models, we have shown that  such variation gives rise
to strong features in the scalar power spectra at scales corresponding
to  those leaving  the  horizon during  the  transition. In  addition,
corresponding features occur in the tensor-to-scalar ratio at the same
scales.  In  comparison to \cite{Adams:2001vc}, these  features do not
exhibit  the  oscillations that  occur  for  step  transitions in  the
inflationary potential  itself.  As shown  in Sec.~\ref{sec:model},
the resulting variation in  the tensor-to-scalar ratio can potentially
alleviate the  tension between recent  measurements by the  Planck and
BICEP2 experiments. A detailed comparison to data will be performed in
future work.

Similar transitions  in the  Planck mass were  studied in  a two-field
model. In  this case  however, sharp features are not observed in the
scalar power  spectrum  and  tensor-to-scalar  ratio, as they are washed out by the conversion of isocurvature modes.

\acknowledgments

This work is partially supported by the Lancaster-Manchester-Sheffield
Consortium for Fundamental Physics  under STFC grant ST/J000418/1. The
work  of PM  is  supported in  part  by the  IPPP  through STFC  grant
ST/G000905/1.   PM would also  like to  acknowledge the  conferment of
visiting  researcher status  at  the University  of  Sheffield. SV  is
supported by an STFC PhD studentship.

\appendix

\section{Field equations: two-field model}
\label{sec:2f}

In  this appendix,  we summarize  the pertinent  background  field and
perturbation equations for the  two-field model with action comprising
Eqs.~\eqref{action} and \eqref{twofieldS}.

The Einstein equation takes the form
\be
  G_{\mu \nu} = \frac{1}{F} \bigg[ T^{(\chi)}_{\mu \nu}
  + \varphi_{,\mu} \varphi_{,\nu}
  - \frac{1}{2} g_{\mu \nu} g^{\alpha \beta}
  \varphi_{,\alpha} \varphi_{,\beta}
  + F_{;\mu\nu}
  - g_{\mu \nu} \Box F
  - g_{\mu \nu} U \bigg]~,
\ee
where
\be
  T^{(\chi)}_{\mu \nu} = - \frac{2}{ \sqrt{-g} }
  \frac{ \delta S^{(\chi)} }{ \delta g^{\mu \nu} }
\ee
is the  energy-momentum tensor of  the field $\chi$. Varying  the full
action  Eq.~\eqref{twofieldS} with  respect to  the two  scalar fields
yields their equations of motion:
\begin{subequations}
\begin{align}
  2 \varpi \Box \varphi\ & =\ F_{,\varphi} g^{\mu\nu}T^{(\chi)}_{\mu\nu}
  - \varpi_{,\varphi} g^{\mu \nu} \varphi_{,\mu} \varphi_{,\nu}
  - 4 F_{,\varphi} U
  + 2F U_{,\varphi}~,
  \\
  \Box \chi \ & =\ V_{,\chi}~.	
\end{align}
\end{subequations}
Lastly, the Friedmann equations take the following forms:
\begin{subequations}
\begin{align}
  H^2 &\ = \ \frac{1}{3F} \bigg[ \frac{1}{2} \dot{\varphi}^2
  + \frac{1}{2} \dot{\chi}^2 + U + V - 3H \dot{F} \bigg]~,
  \\
  -2 \dot{H} &\ =\  \frac{1}{F} \bigg[  \dot{\varphi}^2
  +  \dot{\chi}^2 +  \ddot{F} - H \dot{F}  \bigg]~.
\end{align}
\end{subequations}

The    relevant    perturbation   equations    may    be   found    in
\cite{White:2012ya, Kaiser:2010yu} and are given by
\begin{subequations}
\begin{align}
  \delta \ddot{\varphi} & + \bigg[ 3H
  + \frac{\varpi,_{\varphi}}{\varpi} \dot{\varphi} \bigg]
  \delta \dot{\varphi}
  \nonumber
  \\
  & +  \bigg[ \frac{1}{2}
  \bigg( \frac{F,_{\varphi}}{\varpi} \bigg)_{,\varphi} T^{(\chi)}
  + \frac{1}{2} \bigg( \frac{\varpi,_{\varphi}}{\varpi} \bigg)_{,\varphi}
  \dot{\varphi}^2
  - \frac{1}{2} \bigg( \frac{1}{\varpi} ( 4F,_{\varphi}U
  - 2FU,_{\varphi}  ) \bigg)_{,\varphi} + \frac{k^2}{a^2} \bigg]
  \delta \varphi
  \nonumber \\
  & - ( \dot{\Psi} + 3 \dot{\Phi} ) \dot{\varphi}
  + \frac{1}{\varpi} \bigg[ F_{,\varphi} T^{(\chi)} - 4F_{,\varphi} U
  + 2F U_{,\varphi}  \bigg] \Psi
  \nonumber \\
  & + \frac{1}{2 \varpi} F_{, \varphi} \delta T^{(\chi)} = 0~,	
  \\				
  \delta \ddot{\chi} & +  3H \delta \dot{\chi}
  + \frac{k^2}{a^2} \delta \chi -
  ( \dot{\Psi} + 3 \dot{\Phi}) \dot{\chi}
  + \, 2V,_{\chi} \Psi + V,_{\chi \chi}  \delta \chi = 0~.
\end{align}
\end{subequations}
The rhs's of the perturbed Einstein equations \eqref{Einseq} are given by
\begin{subequations}
\begin{align}
  \delta \rho & =  \frac{1}{F} \bigg[ \dot{\varphi} \delta \dot{\varphi}
  - \dot{\varphi}^2 \Psi + \dot{\chi} \delta \dot{\chi}
  - \dot{\chi}^2 \Psi
  + ( U,_\varphi \delta \varphi + V,_\chi \delta \chi ) 	
  + \, 3 \dot{F} ( \dot{\Phi} + 2H \Psi )
  \nonumber	\\
 & \qquad - 3H ( \delta \dot{F} + H \delta F )
 - \frac{k^2}{a^2} \delta F \bigg]~,	
  \\
  \delta q & =  - \frac{1}{F} \bigg[ \dot{\varphi} \delta \varphi
  + \dot{\chi} \delta \chi + \delta \dot{F}
  - \dot{F} \Psi - H \delta F \bigg]~,
  \\
  \delta p & =  \frac{1}{F} \bigg[ \dot{\varphi} \delta \dot{\varphi}
  - \dot{\varphi}^2 \Psi + \dot{\chi} \delta \dot{\chi}
  - \dot{\chi}^2 \Psi
  - ( U,_\varphi \delta \varphi + V,_\chi \delta \chi ) 		
  - \, p_{\rm tot} \delta F + \delta \ddot{F} + 2H \delta \dot{F}
  \nonumber	\\
  & \qquad - \dot{F} \dot{\Psi} - 2\dot{F} \dot{\Phi}
  - \, 2( \ddot{F} + 2H \dot{F} ) \Psi
  + \frac{k^2}{a^2} \delta F \bigg]~,
\end{align}
\end{subequations}
where $ p_{\mathrm{tot}} $ is the total effective pressure from the two fields.
The latter is defined as
\be
p_{\mathrm{tot}} = \frac{1}{F} \bigg[ \frac{1}{2} \dot{\varphi}^2 
  + \frac{1}{2} \dot{\chi}^2 - U - V + \ddot{F} + 2 H \dot{F} \bigg]~.
\ee
\section{Approximate analytic solution: single-field models}
\label{sec:app1}

\begin{figure}
\begin{center}
  \includegraphics[scale=0.5]{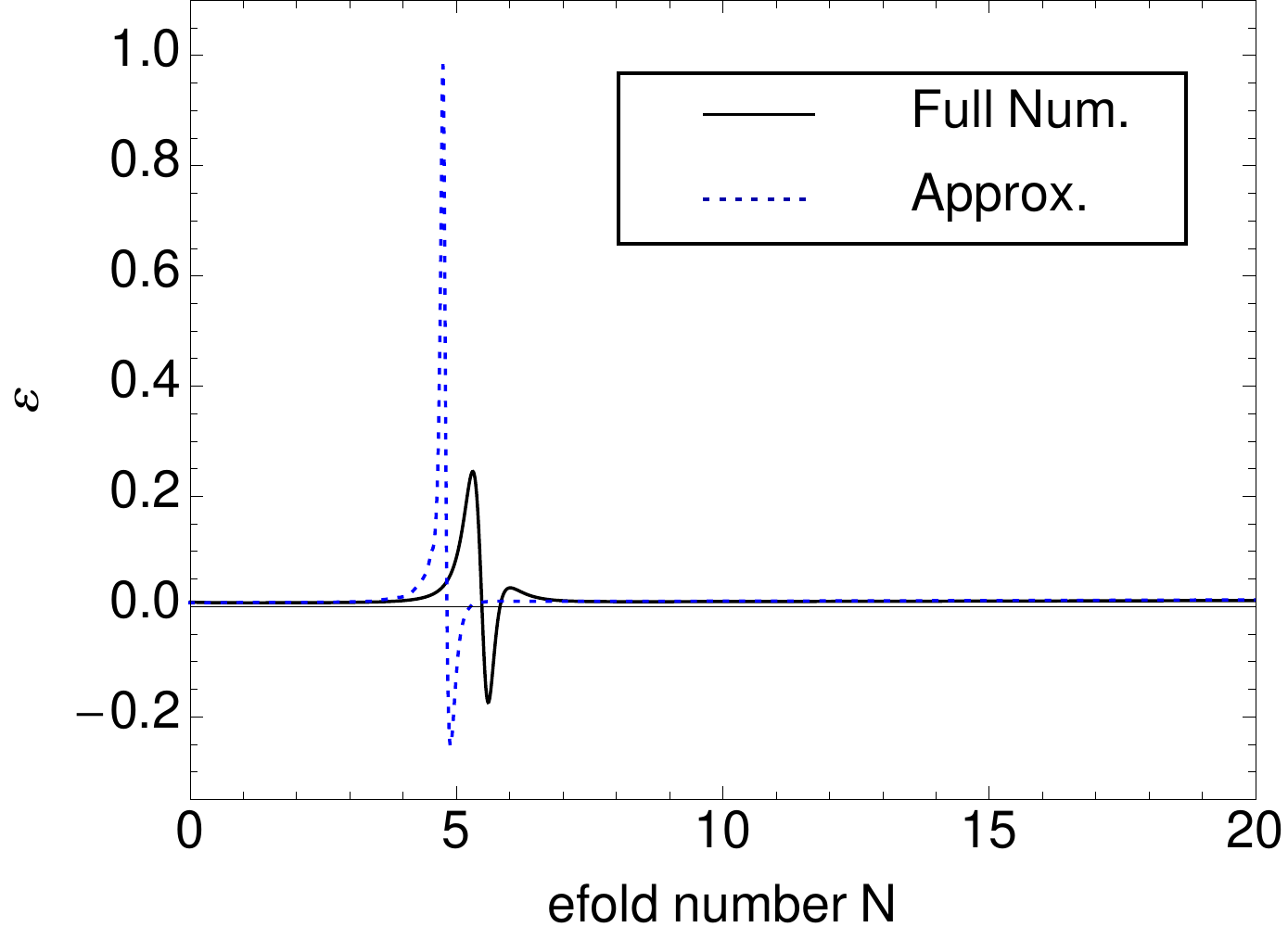}\qquad
  \includegraphics[scale=0.5]{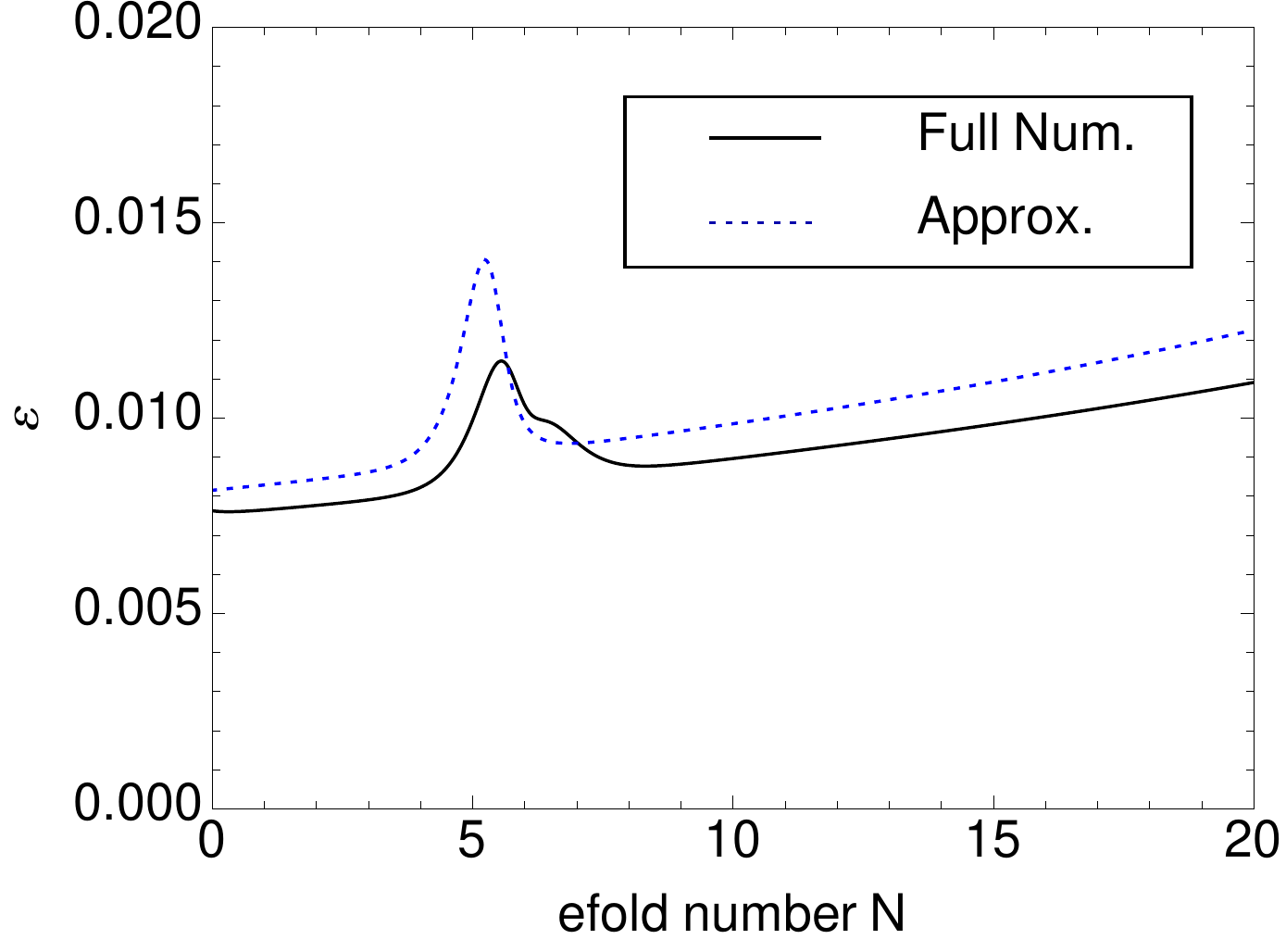}\vspace{2.5em} \\
  \includegraphics[scale=0.5]{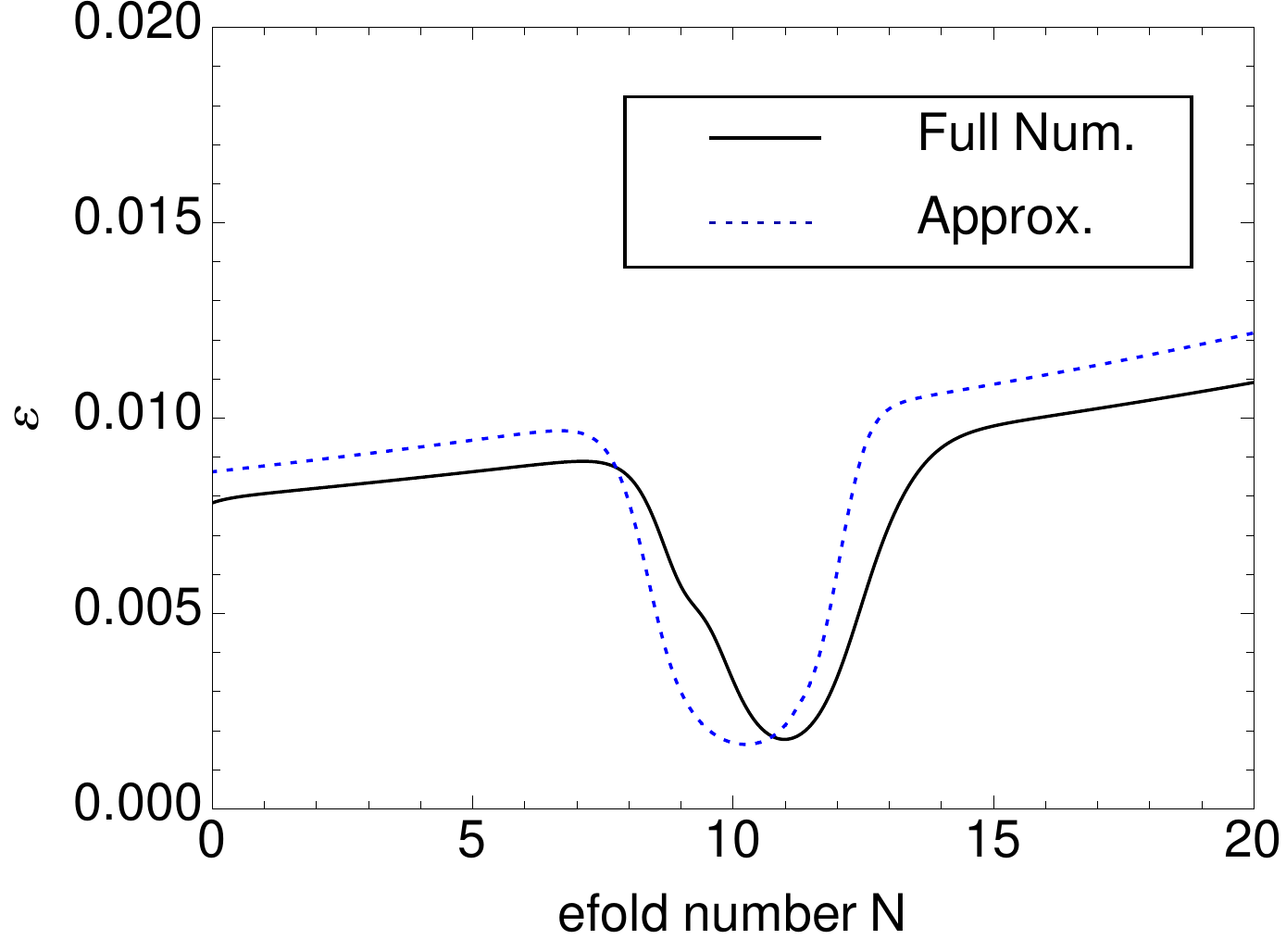}
\end{center}
  \caption{Comparison of  the full numerical  (\emph{solid black}) and
  approximate analytic (\emph{dotted  blue}) solutions for the slow-roll
  parameter  for  benchmark models  1  (\emph{top  left}), 2  (\emph{top
  right})  and (\emph{bottom}) 3.  The $e$-fold  number scaling  for the
  approximate             solution             was            determined
  semi-analytically.\label{compare}}
\end{figure}

\begin{figure}
\begin{center}
  \includegraphics[scale=0.5]{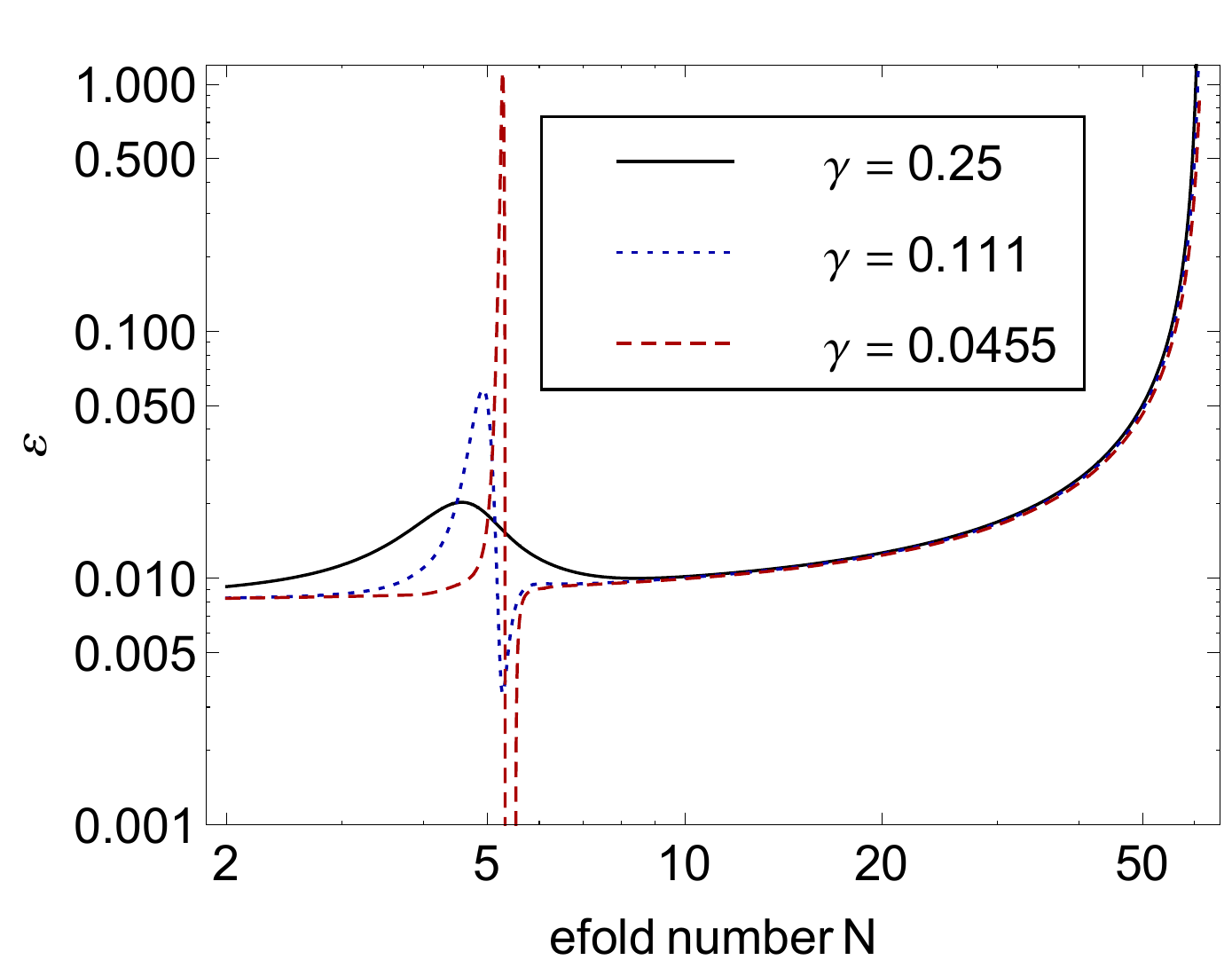}\qquad
  \includegraphics[scale=0.5]{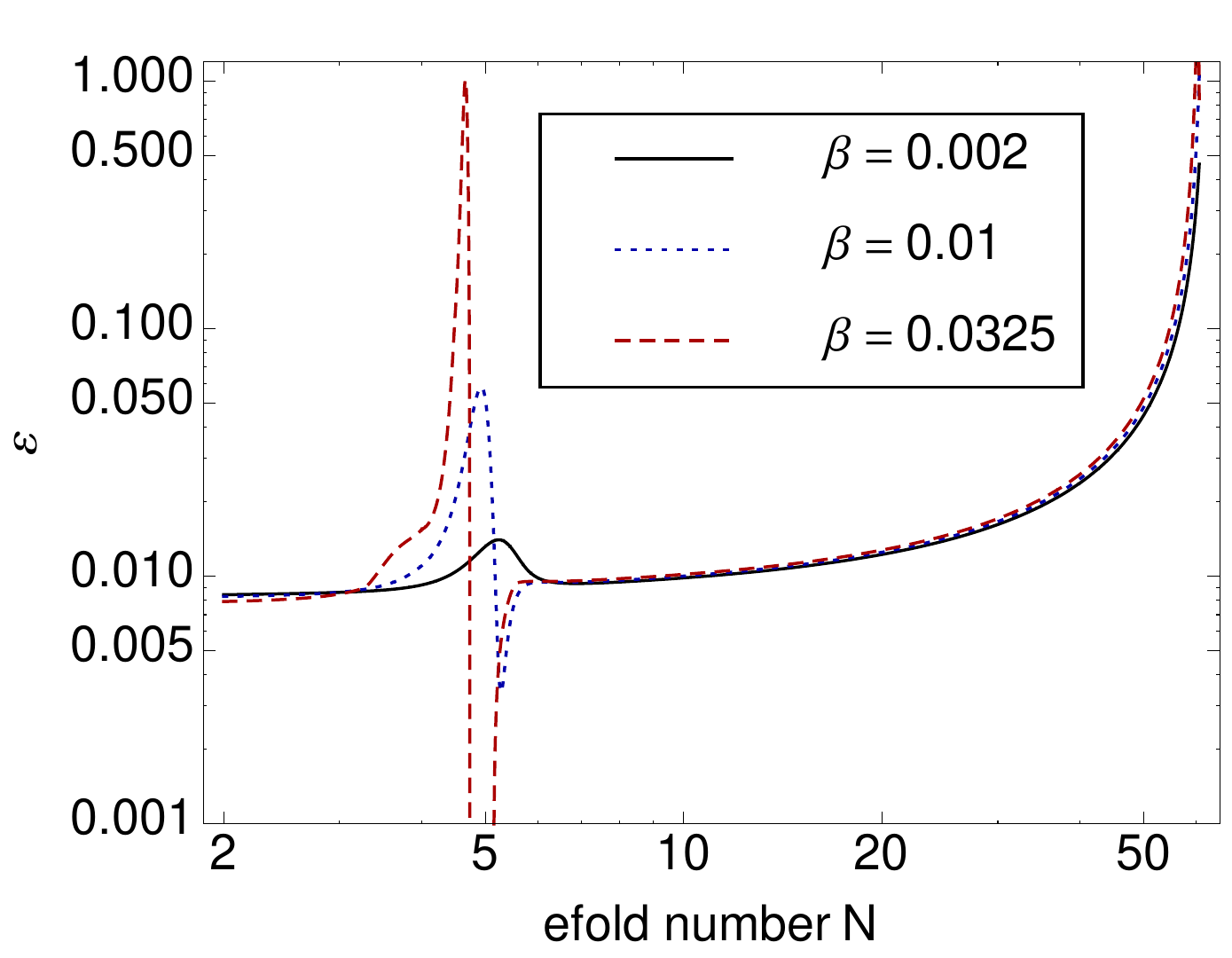}
\end{center}
    \caption{Approximate   evolution   of   the  slow-roll   parameter
    $\varepsilon$  for a  range of  values of  the  parameter $\gamma$
    (\emph{left}, $\beta = 0.01$) and $\beta$ (\emph{right}, $\gamma =
    0.111$). The  remaining model  parameters were those  of benchmark
    model  2: $m_{\varphi}=6.9\times  10^{-6}$, $\varphi_{\ast}=15.49$
    and $\varphi_{\mathrm{ini}}=16.2271$. \label{ranges}}
\end{figure}

In order  to cross-check the  numerical simulations and  to understand
the behaviour of the features in terms of variations of the parameters
$\beta$  and $\gamma$,  it is  illustrative to  obtain  an approximate
analytic  expression for  the phase diagram  of the  Brans-Dicke field
$\varphi$  and the  slow-roll  parameter $\varepsilon$.   By means  of
Eqs.~\eqref{KG} and \eqref{fried}, we may show that $\dot{\varphi}$ is
given by the negative root of
\be
  \dot{\varphi}^2\ =\ \frac{1}{2A}\bigg[-B+\sqrt{B^2+4AC^2}\bigg]\;,
\ee
where
\begin{subequations}
\begin{align}
  A(\varphi) \ & =\ 6\bigg(\frac{\varpi}{F}\bigg)^2F
    -\varpi_{,\varphi}^2+3\big(\varpi^2\big)_{,\varphi}\bigg(\frac{F_{,\varphi}}{F}\bigg)\;,\qquad	\\
  B(\varphi) \ & =\ 12\bigg(\frac{\varpi}{F}\bigg)^2UF
    +2C\bigg(\varpi_{,\varphi}-3\varpi\frac{F_{,\varphi}}{F}\bigg)\;,\qquad	\\
  C(\varphi) \ & =\ 4F_{,\varphi}U-2FU_{,\varphi}
   -2\varpi\ddot{\varphi}\;.
\end{align}
\end{subequations}

In  order  to   find  $\dot\varphi\equiv  \dot{\varphi}(\varphi)$,  we
proceed iteratively under the assumption that $\ddot{\varphi}$ remains
small  in spite of  the transient  features due  to transition  in the
effective Planck mass. In this way, we may approximate
\be
  C \approx C_1 \ =\ 4F_{,\varphi}U-2FU_{,\varphi}
   -2\varpi\ddot{\varphi}_1\;,
\ee
with
\be
  \ddot{\varphi}_1\ =\ \frac{\mathrm{d}}{\mathrm{d} t}\dot{\varphi}_0\ \approx\ \dot{\varphi}_0\frac{\mathrm{d}}{\mathrm{d}\varphi}\dot{\varphi}_0\;,
\ee
where
\be
  \dot{\varphi}_0^2\ =\ \frac{1}{2A}\bigg[-B_0+\sqrt{B_0^2+4AC_0^2}\bigg]\;,
\ee
in which $C$ has been replaced throughout by
\be
  C_0\ =\ 4F_{,\varphi}U-2FU_{,\varphi}\;.
\ee

Figure~\ref{compare} shows a  comparison of the analytic approximation
and full numerical results for  benchmark models 1--3. We see that the
shapes   of   the   features   are  reproduced   by   the   approximate
solution. However,  for the strong  feature in benchmark model  1, the
amplitude of the  analytic approximation does not model  well the full
numerical  solution. Here, we  conclude that  $\ddot{\varphi}/\mpl$ is
not  sufficiently small  for  this  set of  parameters  for the  first-order approximation detailed above to hold.

In  Fig.~\ref{ranges},  we  show  the  evolution  of  the  slow-roll
parameter  $\varepsilon$ determined  using the  analytic approximation
above for a  range of values for the  parameters $\gamma$ and $\beta$.
The  field-dependency  of  the  $e$-fold  number  $N$  was  determined
semi-analytically. These  plots are indicative of  the tuning possible
for  the  shape  of  the   feature  in  the  slow-roll  parameter  and
subsequently   that    occurring   in   the    scalar   power   spectra
$\mathcal{P}_{\zeta}$    as well as   the    tensor-to-scalar    ratio   $r$.
Specifically, smaller values of  $\beta$ and larger values of $\gamma$
lead to sharper  and more violent features in  the slow-roll parameter
and  therefore in  the scalar  power spectra.  We reiterate  that this
first-order  approximation in $\ddot{\varphi}$  only allows comparison
of the  \emph{shape} of  the feature in  the slow-roll  parameter. The
overestimate  of the  amplitude noted  above  can also  be seen  from
Fig.~\ref{ranges}, in  which the duration  of inflation is  of order
$10\%$  shorter  than the  full  numerical solutions,  decreasing
marginally with  increasing amplitude of the feature  in the slow-roll
parameter.


\end{document}